\documentclass[a4paper,12pt]{article}
\usepackage{fullpage}
\usepackage{setspace}
\usepackage{hyperref}
\usepackage[round,sort&compress]{natbib}
\usepackage{tikz-cd}
\usepackage{tikz}
\usepackage{longtable}
\usepackage[latin1]{inputenc}
\usepackage[english]{babel}
\usepackage{amssymb}
\usepackage{amsfonts}
\usepackage{amsmath}
\usepackage{amsthm}
\usepackage{graphicx,epsfig}
\usepackage{bm}
\usepackage{caption}
\usepackage{subcaption}
\usepackage{tabularx}
\usepackage{mathabx}
\newtheorem{definition}{Definition}
\newtheorem{remark}{Remark}
\newtheorem{example}{Example}
\newtheorem{theorem}{Theorem}
\newtheorem{proposition}{Proposition}
\newtheorem{corollary}{Corollary}
\newtheorem*{problem}{Open Problem}

\title{Characterizing the top trading cycles rule for\\housing markets with lexicographic preferences\\when externalities are limited}

\author{Bettina Klaus\thanks{Faculty of Business and Economics, University of Lausanne, Internef 538, 1015 Lausanne, Switzerland; \textit{e-mail}: \href{mailto:Bettina.Klaus@unil.ch}{\tt Bettina.Klaus@unil.ch}. I gratefully acknowledge financial support from the Swiss National Science Foundation (SNSF) through Project 100018$\_$192583. I thank Di Feng, Claudia Meo, Seckin \"{O}zbilen, Jaeok Park, and William Thomson for their valuable comments. I particularly would like to thank two anonymous referees for their excellent feedback and suggestions that helped me to revise and considerably improve this paper.}}

\date{\today}
\begin{document}

\maketitle

\begin{abstract}\noindent We consider a housing market model with limited externalities where agents care both about their own consumption via demand preferences and about the agent who receives their endowment via supply preferences \citep[we extend the associated lexicographic preference domains introduced in][]{KlausMeo2022}. If preferences are demand lexicographic, then our model extends the classical Shapley-Scarf housing market \citep{ShapleyScarfJME1974} with strict preferences model. Our main result is a characterization of the corresponding top trading cycles (TTC) rule by \textsl{individual rationality}, \textsl{pair efficiency}, and \textsl{strategy-proofness} (Theorem~\ref{theorem:paircharacterization}), which extends that of \citet{EkiciTE2023} from classical Shapley-Scarf housing markets with strict preferences to our model. Two further characterizations are immediately obtained by strengthening \textsl{pair efficiency} to either \textsl{Pareto efficiency} or \textsl{pairwise stability} (Corollaries~\ref{corollary:ParetoEfficiency} and \ref{corollary:pairwise stability}). Finally, we show that as soon as we extend the preference domain to include demand lexicographic as well as supply lexicographic preferences (e.g., when preferences are separable), no rule satisfying \textsl{individual rationality}, \textsl{pair efficiency}, and \textsl{strategy-proofness} exists (Theorem~\ref{proposition:impossibility}).\smallskip

\noindent \textbf{Keywords:} Core; characterization; externalities; housing markets; pair efficiency; pairwise stability; top trading cycles rule.\smallskip

\noindent \textbf{JEL codes:} C71, C78, D62, D63.
\end{abstract}
\pagebreak

\section{Introduction}\label{Section:introduction}

In classical \textit{Shapley-Scarf housing markets} each agent is endowed with an indivisible commodity, for instance a house, and wishes to consume exactly one commodity. Agents have complete, reflexive, and transitive preferences over all existing houses and may be better off by trading houses: exchanges do not involve monetary compensations. An outcome, or allocation, for a Shapley-Scarf housing market is a permutation of the endowment allocation.\smallskip

When preferences are strict, one of the best known solution concepts for Shapley-Scarf housing markets is the \textit{strong core}, based on the absence of coalitions that may reallocate their endowments among themselves such that the  members of the blocking coalition are not worse off while at least one member is better off (i.e., no coalition can \textit{weakly block} a strong core allocation).  The strong core for Shapley-Scarf housing markets is always nonempty \citep{ShapleyScarfJME1974} and  coincides with the unique competitive allocation \citep{RothPostlewaiteJME1977}. Using the so-called top trading cycles (TTC) algorithm \citep[due to David Gale, see][]{ShapleyScarfJME1974}, one can easily determine the unique strong core allocation for any Shapley-Scarf housing market with strict preferences.\smallskip

After the above mentioned seminal papers, a number of studies analyze rules for Shapley-Scarf housing markets with strict preferences in view of their normative properties. \citet{RothEL1982} proves that the rule that assigns the unique strong core allocation is \textsl{strategy-proof}, i.e., no agent can benefit from misrepresenting his preferences. Subsequently, \citet{MaIJGT1994} demonstrates that the ``strong core rule'' is the unique rule satisfying \textsl{Pareto efficiency}, \textsl{strategy-proofness}, and \textsl{individual rationality}, i.e., no agent is worse off after trading. For Shapley-Scarf housing markets with strict preferences, the strong core rule equals the top trading cycles (TTC) rule. Therefore, in \citeauthor{MaIJGT1994}'s characterization it is not clear whether the properties characterize the strong core solution or the TTC rule. In a recent paper, \citet{EkiciTE2023} \citep[see also][]{EkiciSethuramanEL2024} shows that in the above characterization result, \textsl{Pareto efficiency} can be replaced by \textsl{pair efficiency}.\smallskip

\citet{KlausMeo2022} build on \citeauthor{ShapleyScarfJME1974}'s classical model by assuming that an agent may not only care about the object he receives but also about the agent who receives his endowment. More specifically, agents may have traditional (strict) ``demand preferences'' over the objects they receive as well as less traditional (strict) ``supply preferences'' over the agents who receive their endowments.  Models with this type of limited externalities fit well with exchanges that are not permanent, i.e., where the endowments are only temporarily exchanged and eventually return to their original owners \citep[see][for a discussion of this and other motivations for the model]{KlausMeo2022}. \citet{AzizLeeAAMAS2020} have introduced the same problem as ``temporary exchange problem.'' General forms of externalities for Shapley-Scarf housing markets have been analyzed before and we again refer to \citet{KlausMeo2022} where the papers of  \citet{MumcuSaglamEB2007}, \citet{HongPark2022}, and \citet{GrazianoMeoYannelis2020} are discussed in more detail.\smallskip

Classical Shapley-Scarf housing market problems with strict preferences can be embedded into the housing market problems with lexicographic preferences that we present here. Our main focus is then on problems where all agents have strict demand lexicographic preferences, i.e., agents care first about the object they receive before considering who receives their endowment (based on possibly weak supply preferences). \citet{KlausMeo2022} consider a somewhat smaller domain of demand lexicographic preferences (based on strict demand and strict supply preferences) and show that the strong core is nonempty and possibly multi-valued. Both these results also hold on our larger demand lexicographic preference domain.\smallskip

Next, we prove that \citeauthor{EkiciSethuramanEL2024}'s (\citeyear{EkiciSethuramanEL2024}) characterization of the TTC rule extends to our more general model: the (corresponding) TTC rule is the unique rule satisfying \textsl{individual rationality}, \textsl{pair efficiency}, and \textsl{strategy-proofness} (Theorem~\ref{theorem:paircharacterization}). Two further characterizations are immediately obtained by strengthening \textsl{pair efficiency} to either \textsl{Pareto efficiency} or \textsl{pairwise stability} (Corollaries~\ref{corollary:ParetoEfficiency} and \ref{corollary:pairwise stability}).
\citet{AzizLeeAAMAS2020} show that the TTC rule satisfies \textsl{individual rationality}, \textsl{Pareto efficiency}, and \textsl{strategy-proofness}.
Our characterization of the TTC rule with \textsl{Pareto efficiency} (Corollary~\ref{corollary:ParetoEfficiency}) complements this result by showing that it is the only rule satisfying these properties.
For Shapley-Scarf housing markets with egocentric preferences (demand lexicographic preferences are egocentric), \citet[][Proposition~4]{HongPark2022} characterize the TTC rule by \textsl{individual rationality}, \textsl{stability}, and \textsl{strategy-proofness}. For our model with demand lexicographic preferences, our characterization of the TTC rule with \textsl{pairwise stability} (Corollary~\ref{corollary:pairwise stability}) complements this result by weakening \textsl{stability} to \textsl{pairwise stability}.\smallskip

We can now distinguish between the previously coinciding solution concepts of the strong core rule and the TTC rule: since for housing markets with demand lexicographic preferences the strong core can be multi-valued, the properties \textsl{individual rationality}, \textsl{pair efficiency} (\textsl{Pareto efficiency} or \textsl{pairwise stability}), and \textsl{strategy-proofness} clearly characterize the TTC rule and \textit{not} the strong core rule (or correspondence). All the above results translate to markets where all agents have supply lexicographic preferences.\smallskip

Finally, we show that as soon as we extend the preference domain to include demand lexicographic as well as supply lexicographic preferences (e.g., when preferences are separable), no rule satisfying \textsl{individual rationality}, \textsl{pair efficiency}, and \textsl{strategy-proofness} exists (Theorem~\ref{proposition:impossibility}).\smallskip

The paper is organized as follows. In Section~\ref{Section:model} we introduce our housing market model with limited externalities, lexicographic preferences, the strong core, as well as rules and their properties. In Section~\ref{subsectionTTCrule}, we introduce the top trading cycles (TTC) rule, show that it produces a strong core allocation that is \textsl{stable} (Proposition~\ref{proposition:TTCstable}), and that it satisfies \textsl{individual rationality}, \textsl{Pareto (pair) efficiency}, \textsl{(pairwise) stability}, and \textsl{strategy-proofness} (Proposition~\ref{proposition:TTCproperties}). In Section~\ref{subsectionTTCcharacterization}, we present three characterizations of the TTC rule on the domain of demand lexicographic preferences (Theorem~\ref{theorem:paircharacterization}, Corollaries~\ref{corollary:ParetoEfficiency} and \ref{corollary:pairwise stability}) and the impossibility result if we extend the preference domain to also include supply lexicographic preferences (Theorem~\ref{proposition:impossibility}). We conclude in Section~\ref{Section:conclusions}.

\section{The model}\label{Section:model}

The following model is very close to the one introduced in \citet{KlausMeo2022}.

\subsection{Housing markets with limited externalities and lexicographic preferences}\label{subsec:basemodel}
We consider an exchange market with indivisibilities formed by $n$ agents and by the same number of indivisible objects, say houses; let $\bm{N}=\{1,\ldots,n\}$ and $\bm{H}=\{h_1,\ldots,h_n\}$ denote the \textbf{set of agents} and \textbf{houses}, respectively. Each agent owns one distinct house when entering the market, desires exactly one house, and has the option to trade the initially owned house in order to get a better one. All exchanges are made with no transfer of money. We assume that \textbf{agent $\bm{i}$ owns house $\bm{h_i}$}; we also refer to $h_i$ as \textbf{agent $\bm{i}$'s endowment}.\medskip

An \textbf{allocation} $\bm{a}$ is an assignment of houses to agents such that each agent receives exactly one house, that is, a bijection $a: N \rightarrow H$. Alternatively, we will denote an allocation $a$ as a vector $a=(a_1,\ldots,a_n)$ with $a_i \in H$ denoting the house assigned to agent $i\in N$ under allocation $a$.   $\bm{\mathcal{A}}$ denotes the \textbf{set of all allocations} and $\bm{h}=(h_1,\ldots,h_n)$ the \textbf{endowment allocation}. Hence, the set of allocations $\mathcal{A}$ is obtained by permuting the set of houses $H$. A nonempty subset $S$ of $N$ is called a \textbf{coalition}. For any coalition $S\subseteq N$ and any allocation $a \in \mathcal{A}$, let $\bm{a(S)} = \{a_i \in H :  i \in S \}$ be the \textbf{set of houses that coalition $\bm{S}$ receives at allocation $\bm{a}$.} The notation $a(i)$ will be used sometimes instead of $a_i$.\medskip

Up to now we have followed the description of a classical \textit{Shapley-Scarf housing market model} as introduced by \cite{ShapleyScarfJME1974}. In contrast with that model, we assume that each agent cares not only about the house he receives but also about the recipient of his own house. That is, preferences capture limited externalities that are modelled as follows.\medskip

Given an allocation $a \in \mathcal{A}$, the \textbf{allotment of agent} $\bm{i}$ is the pair $\bm{(a(i), a^{-1}(h_i))}\in H \times N$, formed by the house $a(i)$ assigned to agent $i$ and the agent who receives agent $i$'s house, i.e., agent $a^{-1}(h_i)$. Note that $a(i)=h_i$ if and only if $a^{-1}(h_i)=i$, i.e., either both elements of \textbf{agent $\bm{i}$'s endowment allotment $\bm{(h_i,i)}$} occur in his allotment or none. $\bm{\mathcal{A}_i}=\left(H\setminus \{h_i\}\times N\setminus\{i\}\right)\cup\{(h_i,i)\}$ denotes the \textbf{set of all the allotments of agent~$\bm{i}$}.\medskip

Each agent $i \in N$ has a preference relation $\succeq_i$ over the set $\mathcal{A}_i$, that is, $\succeq_i$ is a \textit{transitive}, \textit{reflexive}, and \textit{complete} binary relation. As usual, $\succ_i$ and $\sim_i$ denote the asymmetric and symmetric parts of $\succeq_i$, respectively.\footnote{\citet{KlausMeo2022} assume that preferences are \textbf{strict}, i.e.,  $\succeq_i$ is \textit{antisymmetric}. We relax this strictness assumption here.\label{footnote:lexpref}} We denote the general \textbf{domain of preferences for agent~$\bm{i}$} by $\bm{\mathcal{D}_i}$ and the \textbf{set of preference profiles} by $\bm{\mathcal{D}^N}=\mathcal{D}_1\times \ldots \times \mathcal{D}_n$. To simplify notation, we will drop the agent specific lower index from $\mathcal{D}_i$ (respectively, from subdomains of $\mathcal{D}_i$) and simply write $\bm{\mathcal{D}}$.\medskip

To introduce lexicographic preferences, we assume that an agent $i \in N$ may have\vspace{-0.2cm}
\begin{description}
\item a ``\textbf{demand}'' preference relation $\bm{\succeq_i^d}$ over the set $H$ of houses and\vspace{-0.2cm}
\item a ``\textbf{supply}'' preference relation $\bm{\succeq_i^s}$ over the set $N$ of agents.\vspace{-0.2cm}
\end{description}
We denote the set of demand preferences over $H$ and the set of demand preference profiles by $\bm{\mathcal{D}_{\mathrm{d}}}$ and $\bm{\mathcal{D}^N_{\mathrm{d}}}$, respectively; and the set of supply  preferences over $N$ and the set of supply  preference profiles by $\bm{\mathcal{D}_{\mathrm{s}}}$ and $\bm{\mathcal{D}^N_{\mathrm{s}}}$, respectively.
The smaller strict demand and supply preference subdomains are denoted by $\bm{\widetilde{\mathcal{D}}_{\mathrm{d}}}$ and $\bm{\widetilde{\mathcal{D}}_{\mathrm{s}}}$.\vspace{-0.2cm}

\begin{itemize}

\item The domain $\bm{\mathcal{D}_{\mathrm{dlex}}}$ of \textbf{demand lexicographic preferences}: an agent $i \in N$ has demand lexicographic preferences $\succeq_i\,\in \mathcal{D}$ if there exist strict demand preferences $\succeq_i^d\,\in \widetilde{\mathcal{D}}_{\mathrm{d}}$ and (possibly weak) supply preferences $\succeq_i^s\,\in\mathcal{D}_{\mathrm{s}}$ and he primarily cares about the house he receives and only secondarily about who receives his house, i.e.,  for any $(h,j), (h',k)\in \mathcal{A}_i$,\vspace{-0.3cm}
$$(h,j)\succ_i (h',k)\mbox{ if and only if } h\succ^d_i h'\mbox{ or }[h=h'\mbox{ and }j\succ^s_i k].\vspace{-0.3cm}$$
  \item The domain $\bm{\mathcal{D}_{\mathrm{slex}}}$ of \textbf{supply lexicographic preferences}: an agent $i \in N$ has supply lexicographic preferences $\succeq_i\,\in \mathcal{D}$ if there exist strict supply preferences $\succeq_i^s\,\in\widetilde{\mathcal{D}}_{\mathrm{s}}$ and (possibly weak) demand preferences $\succeq_i^d\,\in \mathcal{D}_{\mathrm{d}}$ and he primarily cares about who receives his house and only secondarily about the house he receives, i.e., for any $(h,j), (h',k)\in \mathcal{A}_i$,\vspace{-0.3cm}
$$(h,j)\succ_i (h',k)\mbox{ if and only if } j\succ^s_i k\mbox{ or }[j=k\mbox{ and }h\succ^d_i h'].\vspace{-0.3cm}$$

\item The domain $\bm{\mathcal{D}_{\mathrm{dlex}}\cup\mathcal{D}_{\mathrm{slex}}}$ of \textbf{mixed lexicographic preferences}: agents either have demand lexicographic or supply lexicographic preferences, i.e., some agents first care about the house they receive (then about who receives their house), while others first care about who receives their house (then about the house they receive).\vspace{-0.2cm}
\end{itemize}

Finally, we define the following corresponding smaller strict lexicographic preference domains  $\bm{\widetilde{\mathcal{D}}_{\mathrm{dlex}}}$,  $\bm{\widetilde{\mathcal{D}}_{\mathrm{slex}}}$, $\bm{\widetilde{\mathcal{D}}_{\mathrm{dlex}}\cup\widetilde{\mathcal{D}}_{\mathrm{slex}}}$ by requiring that the underlying demand and supply preferences are both strict. These \textbf{strict preference domains} are the ones considered in \citet{KlausMeo2022}.\medskip

For each agent $i \in N$, a preference relation on the set of allocations $\mathcal{A}$ can be associated with his preferences $\succeq_i$ over $\mathcal{A}_i$. We use the same notation to denote preferences over allotments and allocations, i.e., for any two allocations $a, b \in \mathcal{A}$, $a \succeq_i b \mbox{ if and only if } (a(i),a^{-1}(h_i)) \succeq_i (b(i), b^{-1}(h_i))$.\medskip

\citet{HongPark2022} consider preferences over allocations with externalities as well. Two of their preference domains are the domain of  \textit{hedonic} and the domain of \textit{egocentric} preferences \citep[][Definitions~4 and 8]{HongPark2022}; loosely speaking, an agent's preferences are \textit{hedonic} if he only cares about the trading cycle he belongs to and his preferences are \textit{egocentric} if he primarily cares about the house he receives. \citet[][Assumption~1]{HongPark2022} also assume that no agent is indifferent between any two allocations that assign different houses to him. Assumption~1 is satisfied by all demand lexicographic preferences in $\mathcal{D}_{\mathrm{dlex}}$  but not by all supply lexicographic preferences in $\widetilde{\mathcal{D}}_{\mathrm{slex}}$. The corresponding assumption for supply lexicographic preferences would be that no agent is indifferent between any two allocations that assign his house to different agents.\medskip

For our model, any preference relation on $\mathcal{A}_i$ is hedonic if it satisfies Assumption~1 of \citet{HongPark2022}. Hence, lexicographic preferences in $\mathcal{D}_{\mathrm{dlex}}$ and in $\widetilde{\mathcal{D}}_{\mathrm{slex}}$ are hedonic. Furthermore, demand lexicographic preferences in $\mathcal{D}_{\mathrm{dlex}}$ are egocentric (a similar notion of egocentricity could be defined for supply lexicographic preferences in $\mathcal{D}_{\mathrm{slex}}$ by focusing on the agent that receive a house instead of the house that an agent receives). We discuss \citeauthor{HongPark2022}'s preference domains in relation to ours in more detail in Appendix~\ref{appendix:HongPark}.\medskip

A \textbf{housing market with limited externalities}, or \textbf{market} for short, is now completely described by the triplet $\bm{(N, h, \succeq)}$, where $N$ is the set of agents, $h$ is the endowment allocation, and $\succeq\,\in \mathcal{D}^N$ is a preference profile. Since the set of agents and the endowment allocation are fixed, we often denote a \textbf{market} by its \textbf{preference profile} $\bm{\succeq}$.

\begin{remark}[\textbf{Embedding Shapley-Scarf housing markets into our model}]\label{remark:embedding}\ \\\normalfont Consider the subdomain of lexicographic preferences $\bm{\widehat{\mathcal{D}}_{\mathrm{dlex}}}$ based on strict demand preferences and indifferent supply preferences, i.e., agents only care about the house they receive but they are indifferent to which agent receives their house. Then, a market $(N, h, \succeq)$ with $\succeq\,\in \widehat{\mathcal{D}}_{\mathrm{dlex}}$ corresponds to a Shapley-Scarf housing markets with strict preferences.~\hfill~$\diamond$
\end{remark}

\subsection{Properties of allocations}\label{subsec:AllocationProperties}

We next introduce the strong core, a solution concept that represents the idea of ``stable exchange'' based on the absence of coalitions that can improve their allotments by reallocating their endowments among themselves.

\begin{definition}[\textbf{Strong core allocations}]\ \\\normalfont Let $\succeq\,\in\mathcal{D}^N$ and $a \in \mathcal{A}$. Then, \textbf{coalition $\bm{S}$ weakly blocks allocation $\bm{a}$} if there exists an allocation $b\in \mathcal{A}$ such that\vspace{-0.2cm}
\begin{description}
\item[\textbf{(a)}]at allocation $b$ agents in $S$ reallocate their endowments, i.e., $b(S)=h(S)$, and\vspace{-0.2cm}
\item[\textbf{(b)}] all agents in $S$ are weakly better off with at least one of them being strictly better off, i.e., for all agents $i \in S$, $(b_i,b^{-1}(h_i)) \mathbin{\succeq_i} (a_i,a^{-1}(h_i))$
     and for some agent $ j \in S$, $(b_j,b^{-1}(h_j)) \mathbin{\succ_j} (a_j,a^{-1}(h_j)).$\vspace{-0.2cm}
\end{description}
Allocation $a$ is a \textbf{strong core allocation} if it is not weakly blocked by any coalition. We denote the \textbf{set of strong core allocations} for market $\succeq$ by $\bm{SC(\succeq)}$.\label{weak&strongCore}
\end{definition}

Our definition of the strong core coincides with the definition of the core for hedonic preferences in \citet[][Definition~2]{HongPark2022}. To be more precise, \citet[][page~7]{HongPark2022} explain that ``When an agent has hedonic preferences, he does not care about the allotments of the agents outside of the trading cycle he belongs to. Thus, the $\omega$-core and the $\alpha$-core of a housing market with hedonic preferences coincide.''  \citet[][Theorem~1]{HongPark2022} show the non-emptiness of the strong core for markets with hedonic preferences if a top trading cycles allocation \`{a} la \citet{HongPark2022} exists. The definition of top trading cycles allocations in \citet{HongPark2022} and here, however, differ (see Section~\ref{subsectionTTCrule}) and \citet[][Theorem~1]{HongPark2022} cannot be used to show that for housing markets with demand lexicographic preferences, the strong core is always non-empty. We show that for each $\succeq\,\in \mathcal{D}^N_{\mathrm{dlex}}$, $SC(\succeq)\neq\emptyset$ in Section~\ref{subsectionTTCrule} (Proposition~\ref{proposition:TTCstable}).\footnote{One can establish the non-emptiness of the strong core of a housing market with demand lexicographic preferences based on results in \citet{HongPark2022} as follows. \citet[][Proposition 3]{HongPark2022} shows that the top trading cycles allocation using the associated demand preferences belongs to the $\omega$-core, while for hedonic preferences (and thus for demand lexicographic preferences), the $\omega$-core and the $\alpha$-core coincide (with the strong core).}
Furthermore, in Section~\ref{subsectionTTCrule} (Example~\ref{example:multiSC}) we discuss \citeauthor{KlausMeo2022}'s example of a housing market with strict demand lexicographic preferences and a multi-valued strong core  \citep[][Example~2]{KlausMeo2022}.\medskip

Next, two weaker requirements than being a strong core allocations are the requirements that an allocation can neither be weakly blocked by a single agent $i$, nor by the whole set of agents $N$.

\begin{definition}[\textbf{Individual rationality}]\normalfont
Let $\succeq\,\in\mathcal{D}^N$. Then, allocation $a \in \mathcal{A}$ is \textbf{individually rational} if for all agents $i \in N$, $(a_i,a^{-1}(h_i)) \mathbin{\succeq_i} (h_i,i)$.
\end{definition}

\textsl{Individual rationality} can be interpreted as a voluntary participation requirement since it guarantees that no agent receives an allotment that is worse than his endowment allotment.

\begin{definition}[\textbf{Pareto optimality}]\normalfont An allocation $a\in \mathcal{A}$ is \textbf{Pareto dominated} by allocation $b \in \mathcal{A}$ if for all agents $ i \in N$, $(b_i,b^{-1}(h_i)) \mathbin{\succeq_i} (a_i,a^{-1}(h_i))$  and for some agent $ j \in N$, $(b_j,b^{-1}(h_j)) \mathbin{\succ_j} (a_j,a^{-1}(h_j))$.
An allocation $a\in \mathcal{A}$ is \textbf{Pareto optimal} if it is not Pareto dominated by another allocation.
\end{definition}

For classical Shapley-Scarf housing markets, \citet{EkiciTE2023} weakened \textsl{Pareto efficiency} by requiring that no pair of agents can strictly gain from swapping their assigned houses. Without externalities, if two agents swapped houses to be better off, the obtained allocation would be a Pareto improvement. However, in our model,  the obtained allocation might not only affect the demand preferences of the two agents that swap, it might at the same time impact other agents' supply preferences, possibly making them worse off. Therefore, in order to maintain the spirit of an efficiency property, we'll require that after the swap, all agents are weakly better off.

\begin{definition}[\textbf{Pair efficiency}]\normalfont
Let $\succeq\,\in\mathcal{D}^N$. Then, allocation $a\in \mathcal{A}$ is \textbf{pair efficient} if there exists no pair of agents $i,j\in N$, $i\neq j$, such that allocation $b \in \mathcal{A}$ that is obtained from $a$ by agents $i$ and $j$ swapping houses $a_i$ and $a_j$, is strictly better for both agents, i.e., $(b_i,b^{-1}(h_i)) \mathbin{\succ_i} (a_i,a^{-1}(h_i))$ and $(b_j,b^{-1}(h_j)) \mathbin{\succ_j} (a_j,a^{-1}(h_j))$, and Pareto dominates allocation $a$.
\end{definition}

Finally, another solution concept based on the idea of ``stable exchange'' was introduced by \citet{RothPostlewaiteJME1977}: an allocation is \textit{stable} when no group of agents can reallocate the houses they have obtained such that each agent in the group is strictly better off.

\begin{definition}[\textbf{Stability}]\normalfont
Let $\succeq\,\in\mathcal{D}^N$. Then, allocation $a \in \mathcal{A}$ is \textbf{stable} if there exists no allocation $b\in \mathcal{A}$ and coalition $S$ such that \begin{description}
\item[\textbf{(a')}]at allocation $b$, agents in $S$ reallocate the houses they have obtained at allocation $a$, i.e., $b(S)=a(S)$, while for each $i\in N\setminus S$, $b(i)=a(i)$, and
\item[\textbf{(b')}] all agents in $S$ are strictly better off, i.e., for all agents $i \in S$, $(b_i,b^{-1}(h_i)) \mathbin{\succ_i} (a_i,a^{-1}(h_i))$.
\end{description}
\end{definition}

\citet{RothPostlewaiteJME1977} show that for Shapley-Scarf housing markets without externalities,  \textsl{stability} is equivalent to \textsl{Pareto
efficiency}, and any allocation in the core is \textsl{stable}. However, when there are externalities, \textsl{stability} is a stronger property than \textsl{Pareto
efficiency}, and it is logically independent of the \textsl{(strong) core} \citep[see, for instance,][Definition~3 and the discussion thereafter]{HongPark2022}. Here, we weaken \textsl{stability} to \textit{pairwise stability} by requiring that no two agents $i$ and $j$ can be strictly better off by swapping the houses they have obtained at allocation $a$.\medskip

\begin{definition}[\textbf{Pairwise stability}]\ \\\normalfont
Let $\succeq\,\in\mathcal{D}^N$. Then, allocation and $a \in \mathcal{A}$ is \textbf{pairwise stable} if there exists no pair of agents $i,j\in N$, $i\neq j$, such that allocation $b \in \mathcal{A}$ that is obtained from $a$ by agents $i$ and $j$ swapping houses $a_i$ and $a_j$, is strictly better for both agents, i.e., $(b_i,b^{-1}(h_i)) \mathbin{\succ_i} (a_i,a^{-1}(h_i))$ and $(b_j,b^{-1}(h_j)) \mathbin{\succ_j} (a_j,a^{-1}(h_j))$.
\end{definition}

Note that for Shapley-Scarf housing markets without externalities, \textsl{pairwise stability} and \textsl{pair efficiency} are equivalent. This is not the case in our model. By definition, \textsl{pairwise stability} implies \textsl{pair efficiency}. The following example illustrates that the converse does not hold.

\begin{example}[\textbf{Pair efficiency does not imply pairwise stability}]\label{example:PE-PS}\ \\\normalfont
Let $N=\{1,2,3\}$ and $h=(h_1,h_2,h_3)$. We assume that $\succeq\,\in \widetilde{\mathcal{D}}^N_{\mathrm{dlex}}$ with the following demand and supply preferences.
\begin{center}
\begin{tabular}{m{0.5cm}m{0.2cm}m{0.5cm} m{0.5cm} m{0.5cm}m{0.2cm}m{0.5cm} m{0.5cm} m{0.5cm}m{0.2cm}m{0.5cm}}
\multicolumn{3}{c}{{Agent 1}}&\multicolumn{1}{c}{} & \multicolumn{3}{c}{{Agent 2}}&\multicolumn{1}{c}{} &\multicolumn{3}{c}{{Agent 3}}\\
$\succeq_1^d$ && $\succeq_1^s$  && $\succeq_2^d$ &&  $\succeq_2^s$ &&  $\succeq_3^d$  && $\succeq_3^s$   \\
  \hline
  $h_3$ && $\cdot$  &&   $h_1$ && $\cdot$    & &   $h_2$ && 2  \\
  $h_1$ && $\cdot$  &&   $h_3$ && $\cdot$   & &   $h_3$ && 3   \\
  $h_2$ && $\cdot$  &&   $h_2$ && $\cdot$  &  &   $h_1$ &&  1  \\
\end{tabular}
\end{center}
The empty columns mean that any linear order for $\succeq_1^s$ and $\succeq_2^s$ can be considered.

Consider allocation $a=(h_1,h_3,h_2)$ and the allocation $b=(h_3,h_1,h_2)$ that is obtained from $a$ by agents 1 and 2 swapping the houses they received at $a$. Note that swapping houses makes agents $1$ and $2$ strictly better off, i.e., $(b_1,b^{-1}(h_1)) \mathbin{\succ_1} (a_1,a^{-1}(h_1))$ and $(b_2,b^{-1}(h_2)) \mathbin{\succ_2} (a_2,a^{-1}(h_2))$. Hence, allocation $a$ is not \textsl{pairwise stable}. Note that allocation $b$ is the only allocation that, starting from $a$, makes two agents strictly better off by swapping houses. However, at $b$, agent 3 is made worse off by the swap. Hence, allocation $a$ is \textsl{pair efficient}.\hfill~\qed
\end{example}

\subsection{Rules and their properties}\label{subsec:RuslesAndProperties}
Let $\widetilde{\mathcal{D}}\subseteq\mathcal{D}$ be a generic preference domain.
A \textbf{rule} $\bm{\varphi}:\widetilde{\mathcal{D}}^N\rightarrow\mathcal{A}$ is a function that associates with each market $\succeq$ an allocation $\varphi(\succeq)\in\mathcal{A}$.\medskip

A rule $\varphi$ is\vspace{-0.2cm}
\begin{itemize}
\item \textbf{individually rational} if it only assigns \textsl{individually rational} allocations;\vspace{-0.2cm}
\item \textbf{Pareto efficient} if it only assigns \textsl{Pareto efficient} allocations;\vspace{-0.2cm}
\item \textbf{pair efficient} if it only assigns \textsl{pair efficient} allocations;\vspace{-0.2cm}
\item \textbf{stable} if it only assigns \textsl{stable} allocations;\vspace{-0.2cm}
\item \textbf{pairwise stable} if it only assigns \textsl{pairwise stable} allocations.
\end{itemize}

The well-known non-manipulability property \textsl{strategy-proofness} requires that no agent can ever benefit from misrepresenting his preferences. Let $\succeq\,\in \widetilde{\mathcal{D}}^N$, $i\in N$, and $\widetilde{\succeq}_{i}\,\in \widetilde{\mathcal{D}}$. Then, $(\widetilde{\succeq}_{i}, \succeq_{-i})$ is the preference profile that is obtained from $\succeq$ when agent $i$ changes his preferences from $\succeq_i$ to $\widetilde{\succeq}_{i}$.

\begin{definition}[\textbf{Strategy-proofness}]\normalfont
A rule $\varphi$ is \textbf{strategy-proof} if for each market $\succeq\,\in \widetilde{\mathcal{D}}^N$, each agent $i\in N$, and each preference relation $\widetilde{\succeq}_{i}\in \widetilde{\mathcal{D}}$, \vspace{-0.2cm} $$\varphi_{i}(\succeq)\mathbin{\succeq_{i}} \varphi_{i}(\widetilde{\succeq}_{i},\succeq_{-i}).\vspace{-0.2cm}$$
\end{definition}

\section{The top trading cycles rule}\label{subsectionTTCrule}

We first consider demand lexicographic preferences. Consider a housing market $\succeq\, \in\mathcal{D}^N_{\mathrm{dlex}}$ and its associated demand preferences $\succeq^d$, which we also refer to as the \textbf{associated Shapley-Scarf housing market}. We then define the \textbf{top trading cycles (TTC) allocation for $\bm{\succeq^d}$} using Gale's \textbf{top trading cycles (TTC) algorithm} \citep[][attributed the TTC algorithm to David Gale]{ShapleyScarfJME1974} as follows:\smallskip

\noindent \textbf{Input.} A Shapley-Scarf housing market $\succeq^d\,\in \mathcal{D}^N_{\mathrm{d}}$.\smallskip

\noindent\textbf{Step~1.} Let $N_1:=N$ and $H_1:=H$. We construct a directed graph with the set of nodes $N_1\cup H_1$.
For each agent $i\in N_1$, we add a directed edge to his most preferred house in $H_1$. For each directed edge $(i,h)$,  we say that agent $i$ points to house $h$.
For each house $h\in H_1$, we add a directed edge to its owner.

A \textbf{trading cycle} is a directed cycle in the graph.
Given the finite number of nodes, at least one trading cycle exists. We assign to each agent in a trading cycle the house he points to and remove all trading cycle agents and houses. We define $N_{2}$ to be the set of remaining agents and $H_{2}$ to be the set of remaining houses and, if $N_2\neq\emptyset$, we continue with Step~$2$. Otherwise, we stop.\smallskip

In general, at Step $t$ we have the following:\smallskip

\noindent\textbf{Step~$\bm{t}$.} We construct a directed graph with the set of nodes $N_t\cup H_t$ where $N_t\subseteq N$ is the set of agents that remain after Step~$t-1$ and $H_t\subseteq H$ is the set of houses that remain after Step~$t-1$.
For each agent $i\in N_t$, we add a directed edge to his most preferred house in $H_t$.
For each house $h\in H_t$, we add a directed edge to its owner.

At least one trading cycle exists and we assign to each agent in a trading cycle the house he points to and remove all trading cycle agents and houses. We define $N_{t+1}$ to be the set of remaining agents and $H_{t+1}$ to be the set of remaining houses and, if $N_{t+1}\neq\emptyset$, we continue with Step~$t+1$. Otherwise, we stop.\smallskip

\noindent \textbf{Output.} The TTC algorithm terminates when each agent in $N$ is assigned a house in $H$ (it takes at most $|N|$ steps). We denote the house in $H$ that agent $i\in N$ obtains in the TTC algorithm by $\mathrm{TTC}_{i}(\succeq^d)$ and the final allocation by $\mathrm{TTC}(\succeq^d)$.\medskip

The \textbf{TTC rule} assigns to each market $\succeq\, \in\mathcal{D}^N_{\mathrm{dlex}}$ with associated Shapley-Scarf housing market $\succeq^d\,\in \mathcal{D}^N_{\mathrm{d}}$, the allocation $\mathrm{TTC}(\succeq^d)$, i.e., $\mathrm{TTC}(\succeq):=\mathrm{TTC}(\succeq^d)$. \citet[][Theorem 2]{RothPostlewaiteJME1977} showed that for each Shapley-Scarf housing market $\succeq^d$,\vspace{-0.2cm}
$$SC\left(\succeq^d\right)=\{\mathrm{TTC}\left(\succeq^d\right)\}.\vspace{-0.2cm}$$

We describe how to obtain the TTC algorithm, the TTC allocation, and the TTC rule for supply lexicographic preferences in Appendix~\ref{Appendix:supplypreferences}.\medskip

The TTC allocation for demand lexicographic preferences is a \textsl{strong core} and \textsl{stable} allocation.

\begin{proposition}\label{proposition:TTCstable}Let $\succeq\, \in\mathcal{D}^N_{\mathrm{dlex}}$.\vspace{-0.2cm}
\begin{itemize}
\item[\textbf{\emph{(i)}}] Then, $\mathrm{TTC}(\succeq)\in SC(\succeq)\neq\emptyset$; in particular, $\mathrm{TTC}(\succeq)$ is \textsl{individually rational} and \textsl{Pareto efficient}.\vspace{-0.2cm}
  \item[\textbf{\emph{(ii)}}] Furthermore, $\mathrm{TTC}(\succeq)$ is \textsl{stable}.
\end{itemize}
\end{proposition}

The proof of Proposition~\ref{proposition:TTCstable} (i) is along the lines of the proof of \citet[][Proposion~4]{KlausMeo2022}; however, the domain of demand lexicographic preferences in \citet{KlausMeo2022} is smaller than the domain of demand lexicographic preferences we consider here (see Footnote~\ref{footnote:lexpref}). Alternatively, since demand lexicographic preferences are hedonic and the implied preferences on allocations are egocentric, the result follows from \citet[][Proposition~3]{HongPark2022}. We provide a direct proof for our domain of demand lexicographic preferences $\mathcal{D}^N_{\mathrm{dlex}}$.

\begin{proof}[\textbf{Proof}]Let $(N, h, \succeq)$ be such that $\succeq\,\in\mathcal{D}^N_{\mathrm{dlex}}$ and $(N, h, \succeq^d)$ be the associated Shapley-Scarf housing market. Denote the TTC allocation $\mathrm{TTC}(\succeq)\equiv a$. Hence, ${SC(\succeq^d)}=\{a\}$ and $a$ is \textsl{stable} for Shapley-Scarf housing market $\succeq^d$.\smallskip

\noindent \textbf{(i)} Assume, by contradiction, that $a \not\in  {SC(\succeq)}$. Then, there exist a coalition $S\subseteq N$ and an allocation $b\in\mathcal{A}$ such that\vspace{-0.2cm}
\begin{description}
\item[\textbf{(a)}] $b(S)=h(S)$, and\vspace{-0.2cm}
\item[\textbf{(b)}] for all agents $i \in S$, $(b(i),b^{-1}(h_i))\succeq_i (a(i),a^{-1}(h_i))$ and for some agent $j \in S$, $(b(j),b^{-1}(h_j))\succ_j (a(j),a^{-1}(h_j))$.\vspace{-0.2cm}
\end{description}
Since preferences are demand lexicographic, (b) implies that for all agents $i \in S$, $b(i)\succeq_i^d a(i)$. Let $S_1=\{i \in S: b(i) \succ_i^d a(i)\}$ and $S_2=\{i \in S: b(i) = a(i)\}$. It cannot be the case that $S_2=S$ since that would imply that for all agents $i \in S$, $b(i)=a(i)$ and  $b^{-1}(h_i)=a^{-1}(h_i)$, contradicting (b). Thus, for some agent $j \in S_1$, $b(j)\succ_j^d a(j)$, and for all agents $i \in S=S_1\cup S_2$, $b(i)\succeq_i^d a(i)$.

Hence, in the associated Shapley-Scarf housing market $(N, h, \succeq^d)$, $S$ weakly blocks $a$ through $b$, which contradicts $a \in SC(\succeq^d)$.
$\mathrm{TTC}(\succeq)\in SC(\succeq)\neq\emptyset$ implies that $\mathrm{TTC}(\succeq)$ is \textsl{individually rational} and \textsl{Pareto efficient}.\smallskip

\noindent \textbf{(ii)} Assume, by contradiction, that allocation $a$ is not \textsl{stable}. Then, there exist a coalition $S\subseteq N$ and an allocation $b\in\mathcal{A}$ such that\vspace{-0.2cm}
\begin{description}
\item[\textbf{(a')}] $b(S)=a(S)$ and for each $i\in N\setminus S$, $b(i)=a(i)$, and\vspace{-0.2cm}
\item[\textbf{(b')}] for all agents $i \in S$, $(b_i,b^{-1}(h_i)) \mathbin{\succ_i} (a_i,a^{-1}(h_i))$.\vspace{-0.2cm}
\end{description}
Since preferences are demand lexicographic, (b') implies that for all agents $i \in S$, $b(i)\succeq_i^d a(i)$.  Let $S_1=\{i \in S: b(i) \succ_i^d a(i)\}$ and $S_2=\{i \in S: b(i) = a(i)\}$.
It cannot be the case that $S_2=S$ since that would imply that for all agents $i \in S$, $b(i)=a(i)$ and  $b^{-1}(h_i)=a^{-1}(h_i)$, contradicting (b'). Thus, for some agent $j \in S_1$, $b(j)\succ_j^d a(j)$, and for all agents $i \in S=S_1\cup S_2$, $b(i)\succeq_i^d a(i)$. Furthermore, for all agents $i \in N\setminus S$, $b(i)=a(i)$.
Hence, in the associated Shapley-Scarf housing market $(N, h, \succeq^d)$, $b$ Pareto dominates $a$, which contradicts that TTC allocation $a$ is \textsl{Pareto efficient} for Shapley-Scarf housing market $\succeq^d$.
\end{proof}

The following example shows that the strong core for demand lexicographic preferences can be multi-valued \citep[][Example~2]{KlausMeo2022} and that there can be multiple stable allocations \citep[see also][Example~5]{HongPark2022}.

\begin{example}[\textbf{Multiple strong core and stable allocations}]\label{example:multiSC}\ \\\normalfont
Let $N=\{1,2,3\}$ and $h=(h_1,h_2,h_3)$. We assume that $\succeq\,\in \widetilde{\mathcal{D}}^N_{\mathrm{dlex}}$ with the following demand and supply preferences.
\begin{center}
\begin{tabular}{m{0.5cm}m{0.2cm}m{0.5cm} m{0.5cm} m{0.5cm}m{0.2cm}m{0.5cm} m{0.5cm} m{0.5cm}m{0.2cm}m{0.5cm}}
\multicolumn{3}{c}{{Agent 1}}&\multicolumn{1}{c}{} & \multicolumn{3}{c}{{Agent 2}}&\multicolumn{1}{c}{} &\multicolumn{3}{c}{{Agent 3}}\\
$\succeq_1^d$ && $\succeq_1^s$  && $\succeq_2^d$ &&  $\succeq_2^s$ &&  $\succeq_3^d$  && $\succeq_3^s$   \\
  \hline
  $h_2$ && 3  &&   $h_1$ && 1    & &   $h_2$ && $\cdot$  \\
  $h_3$ && 2  &&   $h_3$ && 3   & &   $h_1$ && $\cdot$    \\
  $h_1$ && 1  &&   $h_2$ && 2  &  &   $h_3$ &&  $\cdot$  \\
\end{tabular}
\end{center}
The empty column means that any linear order for $\succeq_3^s$ can be considered.

The unique strong core for the Shapley-Scarf housing market $\succeq^d$  equals $\mathrm{TTC}(\succeq^d) =(h_2,h_1,h_3)$. Then, $(h_2,h_1,h_3)\in {SC(\succeq)}$.
Next,  $(h_2,h_3,h_1) \in {SC(\succeq)}$ because agent~1 gets his most preferred allotment $(h_2,3)$ and coalition $S=\{2,3\}$ cannot block by swapping their endowments (at allocation $(h_1,h_3,h_2)$ agent~2 would be worse off since $(h_3,1)\mathbin{\succ_2} (h_3,3)$). Note that $(h_2,h_3,h_1) \not\in SC(\succeq)$ because it is weakly blocked by $S=\{2,3\}$ through $(h_1,h_3,h_2)$ (agent~2 receives the same house and agent~3 a better house).  Hence, $(h_2,h_1,h_3),\,(h_2,h_3,h_1) \in {SC(\succeq)}$ but $(h_2,h_1,h_3)\mathbin{\succ_2}(h_2,h_3,h_1)$. Thus, $SC(\succeq)$ is genuinely multi-valued.

TTC allocation $(h_2,h_1,h_3)$ is stable. Allocation $(h_2,h_3,h_1)$ is stable as well because agent~1 gets his most preferred allotment $(h_2,3)$ and if agents 2 and 3 swap the houses $h_3$ and $h_1$ that they obtained at allocation $(h_2,h_3,h_1)$, agent~3 would be worse off ($(h_2,h_3,h_1)\mathbin{\succ_3} (h_2,h_1,h_3)$).\hfill~\qed
\end{example}

\citet[][Theorem~1]{HongPark2022} consider markets with hedonic preferences and state that for any such market, if $a\in\mathcal{A}$ is the TTC allocation (\`{a} la \citeauthor{HongPark2022}), then it is the unique allocation in the core and it is \textsl{stable}. However, note that while our TTC allocation always exists for demand lexicographic preferences, any TTC allocation \`{a} la \citeauthor{HongPark2022} for hedonic preferences must match each agent to his favorite trading cycle. This is not the case for the TTC allocation we obtain in Example~\ref{example:multiSC} \citep[][Example~2]{KlausMeo2022}: at $\mathrm{TTC}(\succeq) =(h_2,h_1,h_3)$ agent 1 trades with agent 2, however, agent 1's favorite trading cycle would be $(1,2,3)$ at which he receives $h_2$ and agent 3 receives $h_1$. Hence, the TTC allocation \`{a} la \citeauthor{HongPark2022} for hedonic preferences does not exist for the problem in Example~\ref{example:multiSC} and no conclusion on the size of the strong core can be drawn. When \citet[][Section~4]{HongPark2022} consider egocentric preferences, their TTC mechanism is similarly defined as our TTC rule.\medskip

The TTC rule satisfies all properties introduced for rules in Section~\ref{subsec:RuslesAndProperties}.

\begin{proposition}\label{proposition:TTCproperties}On $\mathcal{D}^N_{\mathrm{dlex}}$, the TTC rule satisfies \textsl{individual rationality}, \textsl{Pareto (pair) efficiency}, \textsl{(pairwise) stability}, and \textsl{strategy-proofness}.\end{proposition}

\citet{AzizLeeAAMAS2020} also prove that the TTC rule satisfies \textsl{individual rationality}, \textsl{Pareto efficiency}, and \textsl{strategy-proofness} on $\mathcal{D}^N_{\mathrm{dlex}}$. \citet[][Proposition~3]{HongPark2022} implies that the TTC rule for egocentric preferences satisfies \textsl{individual rationality}, \textsl{Pareto efficiency}, and \textsl{group strategy-proofness}.\footnote{A rule satisfies \textbf{group strategy-proofness} on $\mathcal{D}^N_{\mathrm{dlex}}$ if no group of agents can ever benefit by jointly misrepresenting their preferences, i.e., for each market $\succeq$, there is no set of agents $S\subseteq N$ and no preference profile $\widetilde{\succeq}$ such that for each $i\in N\setminus S$, $\widetilde{\succeq}_i=\succeq_i$, for each $j\in S$, $\varphi_{j}(\widetilde{\succeq})\mathbin{\succeq_{i}}\varphi_{j}(\succeq)$, and for some $k\in S$, $\varphi_{k}(\widetilde{\succeq})\mathbin{\succ_{k}}\varphi_{k}(\succeq)$.} We provide a direct proof for our domain of demand lexicographic preferences $\mathcal{D}^N_{\mathrm{dlex}}$.

\begin{proof}[\textbf{Proof}]By Proposition~\ref{proposition:TTCstable}, the TTC rule satisfies \textsl{individual rationality}, \textsl{Pareto (pair) efficiency}, and \textsl{(pairwise) stability} on $\mathcal{D}^N_{\mathrm{dlex}}$.

We next prove that the TTC rule satisfies \textsl{strategy-proofness} on $\mathcal{D}^N_{\mathrm{dlex}}$. Consider a market $\succeq\,\in \mathcal{D}^N_{\mathrm{dlex}}$, an agent $i\in N$, and a preference relation $\widetilde{\succeq}_{i}\in \mathcal{D}_{\mathrm{dlex}}$. Suppose that, by contradiction, agent $i$'s allotment under TTC at $(\widetilde{\succeq}_{i},\succeq_{-i})$ is better than his allotment under TTC at $\succeq$. Then, when moving from $\succeq$ to $(\widetilde{\succeq}_{i},\succeq_{-i})$, either \textbf{(a)} agent $i$ receives a better house or \textbf{(b)} agent $i$ receives the same house and a better agent receives agent $i$'s house.\smallskip

\noindent\textbf{(a)} Since, the TTC rule for Shapley-Scarf housing markets is \textsl{strategy-proof}, the house that agent $i$ receives under TTC at $(\widetilde{\succeq}_{i},\succeq_{-i})$ cannot be better than the one he receives under TTC at $\succeq$.\smallskip

\noindent\textbf{(b)} Since agent $i$ receives the same house under TTC at $\succeq$ and $(\widetilde{\succeq}_{i},\succeq_{-i})$, in order for agent $i$'s house to be assigned to a different agent, some other agent(s) would have to point differently in the execution of the TTC algorithm. However, since agent $i$ is the only agent who changes preferences (and hence possibly his pointing behavior in the TTC algorithm) from $\succeq$ to $(\widetilde{\succeq}_{i},\succeq_{-i})$, all other agents display the same pointing behavior and agent $i$ ends up in the same top trading cycle. Thus, if agent $i$ receives the same house under TTC at $\succeq$ and $(\widetilde{\succeq}_{i},\succeq_{-i})$, then the same agent receives agent $i$'s house under TTC at $\succeq$ and $(\widetilde{\succeq}_{i},\succeq_{-i})$.

To summarize, since agents throughout the TTC algorithm, in line with their demand lexicographic preferences, primarily care about obtaining the best possible house without influencing who receives their house, the TTC rule is \textsl{strategy-proof} for demand lexicographic housing markets.
\end{proof}

\section{Three characterizations and an impossibility result}\label{subsectionTTCcharacterization}

For Shapley-Scarf housing markets with strict preferences, \citet[][Theorem~1]{EkiciTE2023} \citep[see also][Theorem~1]{EkiciSethuramanEL2024} demonstrates that the TTC rule is the unique rule satisfying \textsl{individual rationality}, \textsl{pair efficiency}, and \textsl{strategy-proofness}. We establish a corresponding characterization for the more general model with demand lexicographic preferences (see Remark~\ref{remark:embedding}). A corresponding characterization can symmetrically be derived by changing the roles of houses and agents for the domain of supply lexicographic preferences (see Appendix~\ref{Appendix:supplypreferences} for the definition of the TTC rule on the domain of supply lexicographic preferences).

\begin{theorem}[\textbf{TTC characterization with pair efficiency}]\label{theorem:paircharacterization}\ \\
On $\mathcal{D}^N_{\mathrm{dlex}}$, a rule satisfies \textsl{individual rationality}, \textsl{pair efficiency}, and \textsl{strategy-proofness} if and only if it is the TTC rule.
\end{theorem}

Our proof follows the proof strategy of \citet[][Theorem~1]{EkiciSethuramanEL2024} but it includes additional arguments that are necessary due to the externalities created by agents' supply preferences.

\begin{proof}[\textbf{Proof}]By Proposition~\ref{proposition:TTCproperties}, the TTC rule satisfies \textsl{individual rationality}, \textsl{pair efficiency},  and \textsl{strategy-proofness} on $\mathcal{D}^N_{\mathrm{dlex}}$.

We complete the proof by showing that there is at most one rule satisfying \textsl{individual rationality}, \textsl{pair efficiency}, and \textsl{strategy-proofness} on $\mathcal{D}^N_{\mathrm{dlex}}$. Note that we on purpose closely follow the proof of \citet[][Theorem~1]{EkiciSethuramanEL2024} but that we need to add arguments to deal with agents' supply preferences and to make sure that towards the end of the proof, two agents swapping their assigned houses constitutes a Pareto improvement.
We first adapt their notion of the \textit{size of a preference profile}. Let $\succeq\,\in\mathcal{D}^N_{\mathrm{dlex}}$ with associated demand preference profile $\succeq^d$ and associated supply preference profile $\succeq^s$. Then, at profile $\succeq$, for agent $i\in N$, a \textit{house $h$ is acceptable}  if $h\succeq_i^d h_i$ and an \textit{agent $j$ is acceptable} if $j\succeq_i^s i$.
For profile $\succeq$, its \textit{size}, denoted by $s(\succeq)$, is the total number of houses and agents that agents find acceptable at demand preferences $\succeq^d$ and supply preferences $\succeq^s$. That is, $s(\succeq)=\sum_{i\in N}|\{h\in H:h\text{ is acceptable for agent }i\}|+\sum_{i\in N}|\{j\in N:j\text{ is acceptable for agent }i\}|$.
Note that since each agent finds his own house and himself to be acceptable, $s(\succeq)$ has at least size $2n$.

By way of contradiction, let $\Pi$ and $\Psi$ be two
distinct \textsl{individually rational}, \textsl{pair efficient}, and \textsl{strategy-proof} rules on $\mathcal{D}^N_{\mathrm{dlex}}$.
We call $\succeq\,\in \mathcal{D}^N_{\mathrm{dlex}}$ a conflict profile if $\Pi(\succeq)\neq \Psi(\succeq)$. Since $\Pi\neq \Psi$, there is at least one conflict profile. Among them, let $\succeq$ be one with minimal size. Compare the houses assigned to agents at allocations $\Pi(\succeq)$ and $\Psi(\succeq)$ according to their demand preferences $\succeq^d$. Let $N_\Pi\subseteq N$ be the subset of agents whose assigned houses are strictly better at $\Pi(\succeq)$. Let $N_\Psi\subseteq N$ be the subset of agents whose assigned houses are strictly better at $\Psi(\succeq)$. Then, $\widebar{N}=N\setminus (N_\Pi\cup N_\Psi)$ is the subset of agents whose assigned houses are the same at $\Pi(\succeq)$ and $\Psi(\succeq)$; however, note that this does not necessarily mean that an agent in $\widebar{N}$ is indifferent between $\Pi(\succeq)$ and $\Psi(\succeq)$, because even though the house received remains the same, the agent receiving the endowment might change. Let $a:=\Pi(\succeq)$ and $b:=\Psi(\succeq)$. Since $a$ and $b$ are \textsl{individually rational}, for $i\in N_{\Pi}$, $a_i\succ^d_i b_i\succeq^d_i h_i$, for $i\in N_{\Psi}$, $b_i\succ^d_i a_i\succeq^d_i h_i$, and for $i\in \widebar{N}$, $b_i=a_i\succeq^d_i h_i$.

We will show that for each $i\in N_{\Pi}\cup N_{\Psi}$, $a_i$ and $b_i$ are the only acceptable houses for demand preferences $\succeq^d_i$ and $i$ is the only acceptable agent for supply preferences $\succeq^s_i$.
Suppose that this is not true for some $i\in N_{\Pi}\cup N_{\Psi}$. Without loss of generality, let $i\in N_{\Pi}$. Modify agent $i$'s preferences $\succeq_i$ such that ($i$) at his demand preferences, $a_i$ is ordered first, his endowment $h_i$ comes next, and then come the remaining houses in any order and ($ii$) at his supply preferences, agent $i$ is ordered first\footnote{Note that since $a_i\neq h_i$, $a^{-1}(h_i)\neq i$.} and then come the remaining agents without changing their order among each other. Let the other agents preserve their preferences. Call the resulting profile $\succeq'\,\in \mathcal{D}^N_{\mathrm{dlex}}$ and note that $s(\succeq')<s(\succeq)$. Let $a':=\Pi(\succeq')$. By \textsl{strategy-proofness} of $\Pi$ on $\mathcal{D}^N_{\mathrm{dlex}}$, first $a'_i=a_i$ and then, $a'^{-1}(h_i)=a^{-1}(h_i)$. Hence, $\Pi_i(\succeq')=\Pi_i(\succeq)$. Furthermore,
by \textsl{strategy-proofness} of $\Psi$ on $\mathcal{D}^N_{\mathrm{dlex}}$, $\Psi_i(\succeq')\neq a_i$ and by \textsl{individual rationality},  $\Psi_i(\succeq')=h_i$. Thus, $\Pi_i(\succeq')\neq \Psi_i(\succeq')$. But then $\succeq'$ is a conflict profile with a smaller size than $\succeq$, a contradiction.
By the above arguments, we conclude the following:\vspace{-0.2cm}
\begin{itemize}
\item For each $i\in N_{\Pi}$, at $\succeq_i^d$, $a_i$ is ordered first, followed by his endowment $h_i$, and $b_i=h_i$; at $\succeq_i^s$, $i$ is ordered first.\vspace{-0.2cm}
\item For each $i\in N_{\Psi}$, at $\succeq_i^d$, $b_i$ is ordered first, followed by his endowment $h_i$, and $a_i=h_i$; at $\succeq_i^s$, $i$ is ordered first.\vspace{-0.2cm}
\end{itemize}
Let $i\in\widebar{N}$. Note that $a_i$ cannot be the endowment of an agent in $N_{\Psi}$ because at $\Pi(\succeq)=a$, the agents in $N_{\Psi}$ are assigned their endowments. Also, $b_i$ cannot be the endowment of an agent in $N_{\Pi}$ because at $\Psi(\succeq)=b$, the agents in $N_{\Pi}$ are assigned their endowments. But then, since $a_i=b_i$, we find that at allocations $a$ and $b$, the agents in $\widebar{N}$ are assigned one another's endowments. This implies the following:\vspace{-0.2cm}
\begin{itemize}
\item At $a$, the agents in $N_{\Pi}$ are assigned one another's endowments and at demand preference profile $\succeq^d$, the top choice of an agent $i\in N_{\Pi}$ is the endowment of an agent in $N_{\Pi}$.\vspace{-0.2cm}
\item And at $b$, the agents in $N_{\Psi}$ are assigned one another's endowments and at demand preference profile $\succeq^d$, the top choice of an agent $i\in N_{\Psi}$ is the endowment of an agent in $N_{\Psi}$.\vspace{-0.2cm}
\end{itemize}
By assumption, $N_{\Pi}\cup N_{\Psi}\neq \emptyset$. Without
loss of generality, let $N_{\Pi}\neq \emptyset$. Then, for  $a=\Pi(\succeq)$, there exists a sequence of agents $i_1,i_2,\ldots,i_{k+1}\in N_{\Pi}$, $i_{k+1}=i_1$, such that for each $i_s\in\{i_1,i_2,\ldots,i_{k}\}$, at demand preference profile $\succeq^d$, agent  $i_s$'s top two choices are, in order, $a_{i_s}=h_{i_{s+1}}$ and $h_{i_s}$ [$s+1$ modulo $k$].
Furthermore, for $b=\Psi(\succeq)$, each $i_s\in\{i_1,i_2,\ldots,i_{k}\}$ is assigned his endowment $b_{i_s}=h_{i_s}$, while his top choice is $h_{i_{s+1}}$ [$s+1$ modulo $k$]. Thus, the sequence of agents $i_1,i_2,\ldots,i_{k}$ is such that $k\geq 2$.

We next show that, by \textsl{individual rationality} and \textsl{strategy-proofness} of $\Pi$ and $\Psi$ on $\mathcal{D}^N_{\mathrm{dlex}}$, it is without loss of generality to assume that for each agent $i_s$ in the above sequence, at supply preferences $\succeq^s_{i_s}$, agent $i_s$'s top choice is $i_s$ and for each $i_l,i_{l'}\in \{i_1,\ldots,i_{k}\}\setminus\{i_s\}$, if $l<l'$, then $i_l\succ^s_{i_s}i_{l'}$. Suppose that for some agent $i_s\in\{i_1,i_2,\ldots,i_{k}\}$, supply preferences $\succeq^s_{i_s}$ are not as just described. Then, modify agent $i_s$'s preferences $\succeq_{i_s}$ such that ($i$) his demand preferences remain the same and ($ii$) at his supply preferences, agent $i_s$'s top choice is still $i_s$ and for each $i_l,i_{l'}\in \{i_1,\ldots,i_{k}\}\setminus\{i_s\}$, if $l<l'$, then $i_l\succ^s_{i_s}i_{l'}$. Let the other agents preserve their preferences. Call the resulting profile $\widetilde{\succeq}\,\in \mathcal{D}^N_{\mathrm{dlex}}$. Let $\widetilde{a}:=\Pi(\widetilde{\succeq})$ and $\widetilde{b}:=\Psi(\widetilde{\succeq})$. By \textsl{strategy-proofness} of $\Pi$ on $\mathcal{D}^N_{\mathrm{dlex}}$, $\widetilde{a}_{i_s}=a_{i_s}=h_{i_{s+1}}$ [$s+1$ modulo $k$]. Since $\widetilde{a}_{i_{s+1}}\neq h_{i_{s+1}}$, by \textsl{individual rationality} of $\Pi$ on $\mathcal{D}^N_{\mathrm{dlex}}$, $\widetilde{a}_{i_{s+1}}=a_{i_{s+1}}=h_{i_{s+2}}$ [$s+1,s+2$ modulo $k$]. Sequentially applying \textsl{individual rationality} of $\Pi$ on $\mathcal{D}^N_{\mathrm{dlex}}$ to all agents in the sequence then implies that for each $i_{s'}\in\{i_1,i_2,\ldots,i_{k}\}$, $\widetilde{a}_{i_{s'}}=a_{i_{s'}}=h_{i_{s'+1}}$ [$s'+1$ modulo $k$].
Furthermore,
by \textsl{strategy-proofness} of $\Psi$ on $\mathcal{D}^N_{\mathrm{dlex}}$, $\widetilde{b}_{i_s}\neq a_{i_s}=h_{i_{s+1}}$; then, by \textsl{individual rationality}  of $\Psi$ on $\mathcal{D}^N_{\mathrm{dlex}}$,  $\widetilde{b}_{i_s}=h_{i_s}$. Since $\widetilde{b}_{i_{s-1}}\neq h_{i_{s}}$, by \textsl{individual rationality} of $\Psi$ on $\mathcal{D}^N_{\mathrm{dlex}}$, $\widetilde{b}_{i_{s-1}}=h_{i_{s-1}}$ [$s-1$ modulo $k$]. Sequentially applying \textsl{individual rationality} of $\Psi$ on $\mathcal{D}^N_{\mathrm{dlex}}$ to all agents in the sequence then implies that for each $i_{s'}\in\{i_1,i_2,\ldots,i_{k}\}$, $\widetilde{b}_{i_{s'}}=b_{i_{s'}}=h_{i_{s'}}$. We can thus assume that for each agent $i_s\in \{i_1,i_2,\ldots,i_{k}\}$, at supply preferences $\succeq^s_{i_s}$, agent $i_s$'s top choice is $i_s$ and for each $i_l,i_{l'}\in \{i_1,\ldots,i_{k}\}\setminus\{i_s\}$, if $l<l'$, then $i_l\succ^s_{i_s}i_{l'}$.

Next, at profile $\succeq$, modify agent $i_k$'s preferences $\succeq_{i_k}$ such that ($i$) at his demand preferences $\succeq^d_{i_k}$, his top-ranked houses become, in order, $h_{i_1},h_{i_2},\ldots,h_{i_k}$ and ($ii$) his supply preferences remain the same. Let other agents preserve their preferences. Call the resulting profile $\widebar{\succeq}\,\in \mathcal{D}^N_{\mathrm{dlex}}$. Let $\widebar{b}:=\Psi(\widebar{\succeq})$. Note that \textsl{individual rationality} of $\Psi$ on $\mathcal{D}^N_{\mathrm{dlex}}$ imposes no restrictions on allotment $\widebar{b}_{i_k}$. However, \textsl{strategy-proofness} of $\Psi$ on $\mathcal{D}^N_{\mathrm{dlex}}$ implies that $\widebar{b}_{i_k}\neq h_{i_1}$. Thus, for some $i_s\in\{i_2,i_2,\ldots,i_{k}\}$, we have $\widebar{b}_{i_k}=h_{i_s}$. Since $\widebar{b}$ is  \textsl{individually rational}, $\widebar{b}_{i_{s-1}}=h_{i_{s-1}}$. But then at allocation $\widebar{b}$, according to demand preferences $\widebar{\succeq}^d$, agents  $i_{s-1}$ and $i_k$ prefer one another's assigned houses. Let $\widehat{b}$ be the allocation that is obtained from $\widebar{b}$ when agents $i_{s-1}$ and $i_k$ swap their assigned houses. Thus, agents  $i_{s-1}$ and $i_k$ prefer their allotments at $\widehat{b}$ to their allotments at $\widebar{b}$.

If $i_s=i_k$, then, by \textsl{individual rationality}, at $\widebar{b}$ all agents in $\{i_1,\ldots,i_{k-1}\}$ receive their endowments. Furthermore, $i_{s-1}=i_{k-1}$. Hence, since agents $i_{k-1}$ and $i_k$ swap their own houses, no other agent's allotment changes when moving from $\widebar{b}$ to $\widehat{b}$ and it is a Pareto improvement, which contradicts the supposition that $\Psi$ is \textsl{pair efficient} on $\mathcal{D}^N_{\mathrm{dlex}}$.

If $i_s\neq i_k$, then, by \textsl{individual rationality}, at $\widebar{b}$ all agents in $\{i_s,\ldots,i_{k-1}\}$ receive their top-ranked houses while agents in $\{i_1,\ldots,i_{s-1}\}$ receive their endowments. Hence, the only other agent whose allotment changes when agents $i_{s-1}$ and $i_{k}$ swap their houses is agent $i_s$, who still receives the same house $h_{i_{s+1}}$, but now agent $i_{s-1}$ receives his endowment $h_{i_s}$ instead of agent $i_k$, which is an improvement according to supply preferences $\widebar{\succeq}_{i_s}^s$ (recall that $s-1<k$ implies $i_{s-1}\widebar{\succ}^s_{i_s}i_{k}$). Hence, also in this case, moving from $\widebar{b}$ to $\widehat{b}$ is a Pareto improvement, which contradicts the supposition that $\Psi$ is \textsl{pair efficient} on $\mathcal{D}^N_{\mathrm{dlex}}$.
\end{proof}

Our proof of Theorem~\ref{theorem:paircharacterization} works identically on the smaller domain of  strict lexicographic preference profiles $\widetilde{\mathcal{D}}_{\mathrm{dlex}}^N$. Hence, our characterization of the TTC rule is also valid on $\widetilde{\mathcal{D}}_{\mathrm{dlex}}^N$ instead of $\mathcal{D}_{\mathrm{dlex}}^N$. Note that our proof strategy can be used to extend the characterization result to Shapley-Scarf housing markets with egocentric preferences as considered in \citet[][Proposition~4]{HongPark2022}: basically, the result and proof technique on the smaller preference domain can be used to lift the result up to the larger preference domain (on which all properties are still satisfied); see Theorem~\ref{theorem:paircharacterization2} and its proof in Appendix~\ref{appendix:HongPark}.\medskip

For Shapley-Scarf housing markets with strict preferences, \citet{MaIJGT1994} \citep[see also][Theorem~2]{SvenssonSCW99} demonstrated that the TTC rule is the unique rule satisfying \textsl{individual rationality}, \textsl{Pareto efficiency}, and \textsl{strategy-proofness}. Theorem~\ref{theorem:paircharacterization} implies a corresponding characterization for the more general model with demand lexicographic preferences (see Remark~\ref{remark:embedding}).

\begin{corollary}[\textbf{TTC characterization with Pareto efficiency}]\label{corollary:ParetoEfficiency}\ \\
On $\mathcal{D}^N_{\mathrm{dlex}}$, a rule satisfies \textsl{individual rationality}, \textsl{Pareto efficiency}, and \textsl{strategy-proofness} if and only if it is the TTC rule.
\end{corollary}

Since Corollary~\ref{corollary:ParetoEfficiency} was one of the main results in a previous version of this paper, a direct proof is available in Appendix~\ref{appendix:uniquenessproof}.\medskip

For Shapley-Scarf housing markets with egocentric preferences, \citet[][Proposition~4]{HongPark2022} characterize the TTC rule by \textsl{individual rationality}, \textsl{stability}, and \textsl{strategy-proofness}. Since \textsl{pair efficiency} implies \textsl{pairwise stability}, for our model with demand lexicographic preferences, we obtain a corresponding result with \textsl{pairwise stability} instead of \textsl{stability}.

\begin{corollary}[\textbf{TTC characterization with pairwise stability}]\label{corollary:pairwise stability}\ \\
On $\mathcal{D}^N_{\mathrm{dlex}}$, a rule satisfies \textsl{individual rationality}, \textsl{pairwise stability}, and \textsl{strategy-proofness} if and only if it is the TTC rule.
\end{corollary}

In the following theorem, we obtain an impossibility result when extending the preference domain to include mixed lexicographic preferences, even when restricting attention to strict lexicographic preferences.

\begin{theorem}[\textbf{Impossibility with pair efficiency}]\label{proposition:impossibility}Let $\widetilde{\mathcal{D}}\supseteq  \widetilde{\mathcal{D}}_{\mathrm{dlex}}\cup \widetilde{\mathcal{D}}_{\mathrm{slex}}$ and $|N|\geq 3$. Then, no rule defined on $\widetilde{\mathcal{D}}^N$ satisfies \textsl{individual rationality}, \textsl{pair efficiency}, and \textsl{strategy-proofness}. \end{theorem}

Clearly, the above impossibility result persists when we replace \textsl{pair efficiency} by \textsl{Pareto efficiency} or \textsl{(pairwise) stability}.

\begin{proof}[\textbf{Proof}]Let $\widetilde{\mathcal{D}}\supseteq  \widetilde{\mathcal{D}}_{\mathrm{dlex}}\cup \widetilde{\mathcal{D}}_{\mathrm{slex}}$ and $|N|\geq 3$. Let $\varphi$ be a rule that is defined on $\widetilde{\mathcal{D}}^N$. Assume that $\varphi$ satisfies \textsl{individual rationality}, \textsl{pair efficiency}, and \textsl{strategy-proofness}. We derive a contradiction for $|N|= 3$ that can be straightforwardly extended to $|N| > 3$ by adding agents who find only their own endowment acceptable.\medskip

Let $N=\{1,2,3\}$ and $\succeq=(\succeq_1,\succeq_2,\succeq_3)$ such that $\succeq_1,\,\succeq_2\,\in\widetilde{\mathcal{D}}_{\mathrm{dlex}}$ and $\succeq_3\,\in\widetilde{\mathcal{D}}_{\mathrm{slex}}$ with the following strict demand and supply preferences.
\begin{center}
\begin{tabular}{m{0.5cm}m{0.2cm}m{0.5cm} m{0.5cm} m{0.5cm}m{0.2cm}m{0.5cm} m{0.5cm} m{0.5cm}m{0.2cm}m{0.5cm}}
\multicolumn{3}{c}{{Agent 1}}&\multicolumn{1}{c}{} & \multicolumn{3}{c}{{Agent 2}}&\multicolumn{1}{c}{} &\multicolumn{3}{c}{{Agent 3}}\\
$\succeq_1^d$ && $\succeq_1^s$  && $\succeq_2^d$ &&  $\succeq_2^s$ &&  $\succeq_3^s$  && $\succeq_3^d$\\
  \hline
  $h_2$ && 3  &&   $h_3$ && 1    & &   1 && $h_1$  \\
  $h_3$ && 2  &&   $h_2$ && 3   & &   2 && $h_2$   \\
  $h_1$ && 1  &&   $h_1$ && 2  &  &  3 && $h_3$  \\
\end{tabular}
\end{center}
At $\succeq$ there are only two \textsl{individually rational} and \textsl{pair efficient} allocations, $x=(h_2,h_3,h_1)$ (a trading cycle involving all agents) and $y=(h_3,h_2,h_1)$ (a pairwise trade involving agents 1 and 3).\footnote{Allocations $(h_2,h_1,h_3)$ and $(h_3,h_1,h_2)$ are not \textsl{individually rational}, allocation  $(h_1,h_2,h_3) $ is pair dominated by $(h_1,h_3,h_2)$, and allocation $(h_1,h_3,h_2)$ is pair dominated by $(h_3,h_2,h_1)$.}\smallskip

\noindent \textbf{\textit{Case 1.} }$\varphi(\succeq)=x$.

Note that agent 3 strictly prefers allocation $y$ to allocation $x$. Consider the preference profile $\widetilde{\succeq}$ where agents 1 and 2 report the same preferences as before and agent 3 reports preferences $\widetilde{\succeq}_3\,\in\widetilde{\mathcal{D}}_{\mathrm{slex}}$ with the following strict supply preferences\vspace{-0.2cm}
$$1\ \mathbin{\widetilde{\succ}_3^s}\ 3\ \mathbin{\widetilde{\succ}_3^s}\ 2\vspace{-0.2cm}$$ (while not changing his demand preferences). At $\widetilde{\succeq}$, $y$ is now the only \textsl{individually rational} and \textsl{pair efficient} allocation. Hence, $\varphi(\widetilde{\succeq})=y\succ_3 x=\varphi(\succeq)$; contradicting \textsl{strategy-proofness}.\smallskip

\noindent \textbf{\textit{Case 2.}} $\varphi(\succeq)=y$.

Note that agent 1 strictly prefers allocation $x$ to allocation $y$. Consider the preference profile $\widehat{\succeq}$ where agents 2 and 3 report the same preferences as before and agent 1 reports preferences $\widehat{\succeq}_1\,\in\widetilde{\mathcal{D}}_{\mathrm{dlex}}$ with the following strict demand preferences\vspace{-0.2cm}
$$h_2\ \mathbin{\widehat{\succ}_1^d}\ h_1\ \mathbin{\widehat{\succ}_1^d}\ h_3\vspace{-0.2cm}$$ (while not changing his supply preferences). At $\widehat{\succeq}$, $x$ is now the only \textsl{individually rational} and \textsl{pair efficient} allocation. Hence, $\varphi(\widehat{\succeq})=x\succ_1 y=\varphi(\succeq)$; contradicting \textsl{strategy-proofness}.
\end{proof}

\begin{remark}[\textbf{Preference domains for which Theorem~\ref{proposition:impossibility} holds}]\ \\\normalfont The impossibility result of Theorem~\ref{proposition:impossibility} is established using markets with strict lexicographic preferences. In particular, this implies that our impossibility result is obtained for the domains of mixed lexicographic markets $(\widetilde{\mathcal{D}}_{\mathrm{dlex}}\cup \widetilde{\mathcal{D}}_{\mathrm{slex}})^N$ and $(\mathcal{D}_{\mathrm{dlex}}\cup \mathcal{D}_{\mathrm{slex}})^N$. Furthermore, the impossibility result also holds for various natural strict preference domains that were introduced by \cite{KlausMeo2022}: the domains of strict separable, strict additively separable, and strict preferences. By relaxing the strictness of preferences assumption, the impossibility result then also holds on the larger domains of separable, additively separable, and unrestricted preferences.\hfill~$\diamond$
\label{remark:impossibilitydomains}
\end{remark}

\begin{problem}[\textbf{Maximal preference domain(s)}]\ \\\normalfont As mentioned after the proof of Theorem~\ref{theorem:paircharacterization}, the result and proof technique on the smaller preference domain of demand lexicographic preferences can be used to lift the result up to the larger preference domain of egocentric preferences as considered in \citet{HongPark2022}. Thus, all our characterization results (Theorem~\ref{theorem:paircharacterization} and Corollaries~\ref{corollary:ParetoEfficiency} and \ref{corollary:pairwise stability}) hold for egocentric preferences. It is an open problem to determine a maximal domain for our characterization results.
\end{problem}

\section*{Discussion}

\subsection[S]{Corollary~\ref{corollary:ParetoEfficiency} and \citet[][Theorem~1]{SonmezEco1999}}\label{remark:singletoncores} \citet[][Theorem~1]{SonmezEco1999} proved for so-called generalized indivisible goods allocation problems, which include Shapley-Scarf housing markets, that if there exists a rule $\varphi$ that is \textsl{individually rational}, \textsl{Pareto efficient}, and \textsl{strategy-proof}, then the strong core solution is essentially single-valued (i.e., all agents are indifferent between any pair of allocations in the strong core) and the rule $\varphi$ is a selection of the strong core solution itself.\medskip

For housing markets with lexicographic preferences, Example~\ref{example:multiSC} \citep[][Example~2]{KlausMeo2022} shows that the strong core can be multi-valued. Recall that in Example~\ref{example:multiSC}, $(h_2,h_1,h_3),\,(h_2,h_3,h_1) \in {SC(\succeq)}$ but $(h_2,h_1,h_3)\mathbin{\succ_2}(h_2,h_3,h_1) $. Thus, $SC(\succeq)$ is not essentially single-valued.\footnote{Note that for housing markets with lexicographic preferences, the richness condition on agents' preferences as required in \citet{SonmezEco1999} is violated (see Appendix~\ref{appendix:domainrichness}).}\medskip

We conclude that our model with demand lexicographic preferences is an alternative generalization of Shapley-Scarf housing markets to the one that \citet{SonmezEco1999} considers. In contrast to his generalization, in our model \textsl{individual rationality}, \textsl{Pareto efficiency}, and \textsl{strategy-proofness} are compatible even though the strong core may be genuinely multi-valued (i.e., not essentially single-valued).

\subsection[HP]{Our TTC characterization in comparison with \citet[][Proposition~4]{HongPark2022}}\label{remark:HongPartProposition11}

\citet[][Proposition~4]{HongPark2022} prove that on the domain of egocentric preferences, the TTC rule is the only rule satisfying  \textsl{individual rationality}, \textsl{stability}, and \textsl{strategy-proofness}. \textsl{Stability} implies \textsl{Pareto efficiency}, which in turn implies \textsl{pair efficiency}. Alternatively, \textsl{stability} implies \textsl{pairwise stability}. Hence, our characterizations in  Theorem~\ref{theorem:paircharacterization} and Corollaries~\ref{corollary:ParetoEfficiency} and \ref{corollary:pairwise stability} use weaker properties to characterize the TTC rule. On the other hand, we define the TTC rule on a smaller preference domain than \citet{HongPark2022} do. Thus, one assumption in our characterization results is stronger (we consider a smaller preference domain) while another is weaker (we use \textsl{pair efficiency}, \textsl{Pareto efficiency}, or \textsl{pairwise stability} instead of \textsl{stability}) than in \citet[][Proposition~4]{HongPark2022}.\footnote{About a previous version of this paper, \citet[][page~3]{HongPark2022} write ``Compared to our characterization of the TTC mechanism on the domain of egocentric preferences,
her characterization reveals that the weaker property of Pareto efficiency than stability can be used when the preference domain
is restricted to that of demand-lexicographic preferences.''}

The main characterization result of a previous version of this paper with \textsl{Pareto efficiency} (Corollary~\ref{corollary:ParetoEfficiency}) is  logically independent of that of \citet[][Proposition~4]{HongPark2022} (and it was independently obtained). In the meantime, we strengthened our main result (Theorem~\ref{theorem:paircharacterization}) by weakening \textsl{Pareto efficiency} to \textsl{pair efficiency}. Furthermore, we show how the characterization result on the smaller domain of demand lexicographic preferences can be used as a ``stepping stone'' to prove a corresponding characterization result on the larger domain of egocentric preferences (Theorem~\ref{theorem:paircharacterization2} in Appendix~\ref{appendix:HongPark}). It is an \textbf{open problem} if a direct proof of that characterization result without explicit use of demand lexicographic preferences exists.

\section{Conclusion}\label{Section:conclusions}

On the domain of demand lexicographic preferences, the top trading cycles (TTC) rule, is the unique rule satisfying \textsl{individual rationality}, \textsl{pair (Pareto) efficiency}, and \textsl{strategy-proofness}.\medskip

For classical Shapley-Scarf housing markets (with strict preferences), the TTC rule is also characterized by these properties and the TTC allocation equals the unique strong core allocation. This gives rise to two interpretations of this characterization result for Shapley-Scarf housing markets:\vspace{-0.2cm}
\begin{itemize}
\item[(a)]\textsl{individual rationality}, \textsl{pair (Pareto) efficiency}, and \textsl{strategy-proofness} characterize the TTC rule or\vspace{-0.2cm}
\item[(b)]\textsl{individual rationality}, \textsl{pair (Pareto) efficiency}, and \textsl{strategy-proofness} characterize the rule that assigns the strong core solution.
\end{itemize}

We consider a more general model with demand lexicographic preferences that contains the classical Shapley-Scarf housing markets (see Remark~\ref{remark:embedding}) and show that in our model, the properties \textsl{individual rationality}, \textsl{pair (Pareto) efficiency}, and \textsl{strategy-proofness} characterize (a) the TTC rule but \textit{not} (b) the strong core (which can now be multi-valued).\vspace{-0.4cm}

$$\varphi\mbox{ is the TTC rule }$$
$$\quad\quad\quad\quad\quad\quad\quad\quad\quad\quad\quad\!\!\!\!\!\!\!\Big\Updownarrow\quad \mbox{Theorem~\ref{theorem:paircharacterization} /  Corollary~\ref{corollary:ParetoEfficiency}} $$
$$\varphi\mbox{ satisfies \textsl{individual rationality}, \textsl{pair (Pareto) efficiency}, and \textsl{strategy-proofness}}$$
$$\quad\quad\quad\quad\quad\Big\Uparrow\!\!\!\!\!\!\!\diagup\Big\Downarrow\quad \mbox{Proposition~\ref{proposition:TTCstable}}$$
$$\varphi\mbox{ assigns a strong core allocation}$$

Hence, Theorem~\ref{theorem:paircharacterization} (Corollary~\ref{corollary:ParetoEfficiency}) sheds some new light on a classical characterization result for Shapley-Scarf housing markets with strict preferences: the properties \textsl{individual rationality}, \textsl{pair (Pareto) efficiency}, and \textsl{strategy-proofness} may primarily pin down the TTC rule and only secondarily (or coincidentally) induce strong core selection.

\newcommand{\noopsort}[1]{} \newcommand{\printfirst}[2]{#1}
  \newcommand{\singleletter}[1]{#1} \newcommand{\switchargs}[2]{#2#1}

\begin{appendix}
\section*{Appendix}

\section[HP]{Preference domains in \citet{HongPark2022}}\label{appendix:HongPark}

\citet{HongPark2022} consider preferences over allocations with externalities as well. They first require Assumption~1:\vspace{-0.2cm}
$$\mbox{for each agent }i,\ a\sim_i b\mbox{ implies }a(i)=b(i).\vspace{-0.2cm}$$
For our model, for strict preferences, since $a\sim_i b$ implies $(a(i),a^{-1}(h_i)) = (b(i), b^{-1}(h_i))$, the above assumption is satisfied by all strict preference domains we consider, e.g., for $\widetilde{\mathcal{D}}_{\mathrm{dlex}}$,  $\widetilde{\mathcal{D}}_{\mathrm{slex}}$, $\widetilde{\mathcal{D}}_{\mathrm{dlex}}\cup\widetilde{\mathcal{D}}_{\mathrm{slex}}$.
Furthermore, demand lexicographic preferences in $\mathcal{D}_{\mathrm{dlex}}$ satisfy Assumption~1 while supply lexicographic preferences in $\mathcal{D}_{\mathrm{slex}}$ may not satisfy Assumption~1.\footnote{\citet{HongPark2022} define various classes of preferences focusing on preferences over allocations that satisfy Assumption~1 \citep[see][Figure~1]{HongPark2022}.}\medskip

\noindent Next, \citet{HongPark2022} define agent $i$'s preferences $\unrhd_i$ over $\mathcal{A}$  as\vspace{-0.2cm}
\begin{itemize}
\item \textit{hedonic} if each agent just cares about his own \textit{trading cycle};\footnote{For each allocation $a\in\mathcal{A}$, the set $N$ of agents can be partitioned into trading cycles: a trading cycle is a sequence of agents $(j_0, j_1, \ldots , j_{K-1})$ such that for each $i = 0, 1, \ldots , K-1$, $a(j_i) = h_{j_{i+1}}$ (where indices are modulo K).} that is,  for all allocations $a,b \in \mathcal{A}$ such that $\mathcal{S}_i^{a,h}=\mathcal{S}_i^{b,h}$, where $\mathcal{S}_i^{a,h}$ and $\mathcal{S}_i^{b,h}$ are agent $i$'s trading cycles at allocations $a$ and $b$, respectively, we have $a \sim_i b$;\vspace{-0.2cm}
\item  \textit{egocentric} if each agent is primarily interested in the house (or, the \textit{allotment}, according to Hong and Park's terminology) that he receives; formally, for all $a, b \in \mathcal{A}$ with $a(i) \neq b(i)$, it holds that

$a \rhd_i b   \mbox{ implies for all } a',b' \in \mathcal{A}\mbox{ with  }[a'(i)=a(i) \mbox{ and } b'(i)=b(i)]\mbox{ that } a' \rhd_i b'.$

Hence, if preferences $\unrhd_i $ are egocentric, then there exist associated demand preferences $\succeq^d_i$ such that for all $a,b \in \mathcal{A}$, $a(i)\succ^d_i b(i)$ implies $a \rhd_i b$.\vspace{-0.2cm}
\end{itemize}

For our model, since for any two allocations $a,b\in \mathcal{A}$, $\mathcal{S}_i^{a,h}=\mathcal{S}_i^{b,h}$ implies $(a(i),a^{-1}(h_i)) = (b(i), b^{-1}(h_i))$, a preference relation $\unrhd_i$ over allocations that is derived from strict preferences $\succeq_i\,\in \mathcal{D}$ over allotments satisfies the requirement to be hedonic. Thus, throughout this paper, preferences over allocations are hedonic.\smallskip

Preferences $\unrhd$ over allocations that are induced by demand lexicographic preferences $\succeq \in\,\mathcal{D}_{\mathrm{dlex}}$ are also egocentric. On the other hand, our domain of supply lexicographic preferences may induce preferences over allocations that do not satisfy the requirement to be egocentric \citep[see][Remark~1, for an example]{KlausMeo2022}.\medskip

Our main result (Theorem~\ref{theorem:paircharacterization}) also holds on the domain of egocentric preferences.

\begin{theorem}[\textbf{TTC characterization for egocentric preferences}]\label{theorem:paircharacterization2}\ \\
A rule on the domain of egocentric preferences satisfies \textsl{individual rationality}, \textsl{pair efficiency}, and \textsl{strategy-proofness} if and only if it is the TTC rule.
\end{theorem}
\begin{proof}[\textbf{Proof}]By \citet[][Proposition~3]{HongPark2022}, the TTC rule satisfies \textsl{individual rationality}, \textsl{pair (Pareto) efficiency}, and \textsl{(group) strategy-proofness}  for egocentric preferences.

We complete the proof by showing that there is at most one rule satisfying \textsl{individual rationality}, \textsl{pair efficiency}, and \textsl{strategy-proofness} for egocentric preferences.

By way of contradiction, let $\Pi$ and $\Psi$ be two
distinct \textsl{individually rational}, \textsl{pair efficient}, and \textsl{strategy-proof} rules for egocentric preferences.
We call an egocentric preference profile $\unrhd$ a conflict profile if $\Pi(\unrhd)\neq \Psi(\unrhd)$. Since $\Pi\neq \Psi$, there is at least one egocentric conflict profile $\unrhd$. As in the proof of Theorem~\ref{theorem:paircharacterization}, compare the houses assigned to agents at allocations $\Pi(\unrhd)$ and $\Psi(\unrhd)$ according to their associated demand preferences $\succeq^d$, let $N_\Pi\subseteq N$ be the subset of agents whose assigned houses are strictly better at $\Pi(\unrhd)$, let $N_\Psi\subseteq N$ be the subset of agents whose assigned houses are strictly better at $\Psi(\unrhd)$, and let $\widebar{N}=N\setminus (N_\Pi\cup N_\Psi)$ be the subset of agents whose assigned houses are the same at $\Pi(\unrhd)$ and $\Psi(\unrhd)$. Let $a:=\Pi(\unrhd)$ and $b:=\Psi(\unrhd)$. Since $a$ and $b$ are \textsl{individually rational}, for $i\in N_{\Pi}$, $a_i\succ^d_i b_i\succeq^d_i h_i$, for $i\in N_{\Psi}$, $b_i\succ^d_i a_i\succeq^d_i h_i$, and for $i\in \widebar{N}$, $b_i=a_i\succeq^d_i h_i$.

We next show that, by \textsl{individual rationality} and \textsl{strategy-proofness} of $\Pi$ and $\Psi$ for egocentric preferences, it is without loss of generality to assume that for each agent $i\in N_\Pi\cup N_\Psi$, $\unrhd_i\in \mathcal{D}_{\mathrm{dlex}}$.

Suppose that this is not true for some $i\in N_{\Pi}\cup N_{\Psi}$. Without loss of generality, let $i\in N_{\Pi}$. Modify agent $i$'s preferences $\unrhd_i$ such that they are in  $\mathcal{D}_{\mathrm{dlex}}$ and ($i$) his demand preferences $\succeq^d_i$ remain the same and ($ii$) at his supply preferences, agent $i$ is ordered first and then come the remaining agents in an arbitrary order. Let the other agents preserve their preferences. Call the resulting profile $\unrhd'$. Let $a':=\Pi(\unrhd')$. By \textsl{strategy-proofness} of $\Pi$ for egocentric preferences, $a'_i=a_i$. Hence, $\Pi_i(\unrhd')=\Pi_i(\unrhd)$. Furthermore,
by \textsl{strategy-proofness} of $\Psi$ for egocentric preferences, $\Psi_i(\unrhd')\neq a_i$. Thus, $\Pi_i(\unrhd')\neq \Psi_i(\unrhd')$. Hence, $\unrhd'$ is a conflict profile again. We can now repeat the transformation of egocentric preferences into demand lexicographic preferences for another agent who receives different houses $\Pi_i(\unrhd')$ and $\Psi_i(\unrhd')$. Since the set of agents is finite, we finally end up with a conflict profile at which all agents receiving different houses via the two rules have demand lexicographic preferences.

Note that in the proof of Theorem~\ref{theorem:paircharacterization}, the definition of size and minimality of a preference profile can be restricted to the set of agent who receives different houses under rule $\Pi$ and rule $\Psi$. With that adjusted size and minimality definition, the proof now proceeds exactly as that of Theorem~\ref{theorem:paircharacterization}.
\end{proof}

\section{TTC for supply lexicographic preferences}\label{Appendix:supplypreferences}

We now  consider supply lexicographic preferences. Consider a housing market $\succeq\,\in\mathcal{D}^N_{\mathrm{slex}}$ and its associated supply preference profile $\succeq^s$. We then define the \textbf{top trading cycles (TTC) allocation for $\bm{\succeq^s}$} by adapting the \textbf{top trading cycles (TTC) algorithm} as follows:\medskip

\noindent \textbf{Input.} A supply lexicographic preference profile $\succeq^s\,\in \mathcal{D}^N_{\mathrm{s}}$.\smallskip

\noindent\textbf{Step~1.} Let $N_1:=N$ and $H_1:=H$. We construct a directed graph with the set of nodes $N_1\cup H_1$.
For each agent $i\in N_1$, we add a directed edge to his house $h_i$.
For each house $h\in H_1$, we add a directed edge to the agent that its owner most prefers in $N_1$. For each directed edge $(h,i)$,  we say that house $h$ points to agent $i$.

A \textbf{trading cycle} is a directed cycle in the graph.
Given the finite number of nodes, at least one trading cycle exists. We assign to each house $h$ in a trading cycle the agent $i$ it points to, i.e., agent $i$ receives house $h$, and remove all trading cycle agents and houses. We define $N_{2}$ to be the set of remaining agents and $H_{2}$ to be the set of remaining houses and, if $N_2\neq\emptyset$, we continue with Step~$2$. Otherwise, we stop.\smallskip

In general, at Step $t$ we have the following:\smallskip

\noindent\textbf{Step~$\bm{t}$.} We construct a directed graph with the set of nodes $N_t\cup H_t$ where $N_t\subseteq N$ is the set of agents that remain after Step~$t-1$ and $H_t\subseteq H$ is the set of houses that remain after Step~$t-1$.
For each agent $i\in N_t$, we add a directed edge to his house $h_i$.
For each house $h\in H_t$, we add a directed edge to the agent that its owner most prefers in $N_t$.

At least one trading cycle exists and we assign to each house $h$ in a trading cycle the agent $i$ it points to and remove all trading cycle agents and houses. We define $N_{t+1}$ to be the set of remaining agents and $H_{t+1}$ to be the set of remaining houses and, if $N_{t+1}\neq\emptyset$, we continue with Step~$t+1$. Otherwise, we stop.\smallskip

\noindent \textbf{Output.} The TTC algorithm terminates when each agent in $N$ is assigned a house in $H$ (it takes at most $|N|$ steps). We denote the house in $H$ that agent $i\in N$ obtains in the TTC algorithm by $\mathrm{TTC}_{i}(\succeq^s)$ and the final allocation by $\mathrm{TTC}(\succeq^s)$.\medskip

The \textbf{TTC rule} assigns to each market $\succeq\, \in\mathcal{D}^N_{\mathrm{slex}}$ with associated supply preference profile $\succeq^s\,\in \mathcal{D}^N_{\mathrm{s}}$, the allocation $\mathrm{TTC}(\succeq^s)$, i.e., $\mathrm{TTC}(\succeq):=\mathrm{TTC}(\succeq^s)$.

\section{Direct proof of Corollary~\ref{corollary:ParetoEfficiency}}\label{appendix:uniquenessproof}

\begin{proof}[\textbf{Proof of Corollary~\ref{corollary:ParetoEfficiency}}]\ \\By Proposition~\ref{proposition:TTCproperties}, the TTC rule satisfies \textsl{individual rationality}, \textsl{Pareto efficiency}, and \textsl{strategy-proofness} on $\mathcal{D}^N_{\mathrm{dlex}}$.\smallskip

Let $\varphi$ be a rule satisfying \textsl{individual rationality}, \textsl{Pareto efficiency}, and \textsl{strategy-proofness} on $\mathcal{D}^N_{\mathrm{dlex}}$. Let $\succeq\,\in\mathcal{D}^N_{\mathrm{dlex}}$ with associated Shapley-Scarf housing market $\succeq^d\,\in \mathcal{D}^N_{\mathrm{d}}$. We show that $\mathrm{TTC}(\succeq)=\varphi(\succeq)$ following the steps of the TTC algorithm. \smallskip

The intuition for the first proof step is as follows. Consider a trading cycle that forms in the first step of the TTC algorithm for  $\succeq$. Note that an agent who points at his own house will receive it due to \textsl{individual rationality} of $\varphi$. For larger trading cycles, agents in the trading cycle receive their most preferred houses under TTC, which are different from their own houses. The proof that agents in that trading cycle receive their TTC houses under $\varphi$ as well proceeds as follows. By \textsl{individual rationality} of $\varphi$, trading cycle agents receive a house that is at least as preferred as their own house. Now, imagine that demand preferences for trading cycle agents were such that each of them ranks their own house just below their TTC house. Then, preferences being demand lexicographic, by \textsl{individual rationality} of $\varphi$, each trading cycle agent either receives their TTC or their own house. However, then the only two feasible allotments under $\varphi$ for trading cycle agents are either the endowment allotments or the TTC allotments. By \textsl{Pareto efficiency},  trading cycle agents then all must receive their TTC house under $\varphi$. \smallskip

\noindent \textbf{TTC algorithm Step~1 trading cycles}

Consider a trading cycle that forms in the first step of the TTC algorithm for
$\succeq$. If the trading cycle consists of only one agent, then that agent, according to his demand preferences, most prefers his own house and, by \textsl{individual rationality} of $\varphi$, receives it; such an agent receives his TTC allotment.

Hence, consider a first step trading cycle consisting of agents $i_0,\ldots,i_{K}$, $K\geq 1$, and houses $h_{i_0},\ldots,h_{i_{K}}$.  Note that each agent $i_k\in \{i_0,\ldots,i_{K}\}$, according to his demand preferences $\succeq_{i_k}^d$, prefers house $h_{i_{k+1}}$ most among houses in $H$.

For every $i_k\in \{i_0,\ldots,i_{K}\}$  we define preferences $\widetilde{\succeq}_{i_k}\,\in \mathcal{D}_{\mathrm{dlex}}$  based on new demand preferences $\widetilde{\succeq}^d$  and the original supply preferences ${\succeq}^s$ such that\vspace{-0.2cm}
\begin{itemize}
\item $h_{i_{k+1}}\mathbin{\widetilde{\succ}^d_{i_{k}}}h_{i_k} \mathbin{\widetilde{\succ}^d_{i_k}}\ldots$  (modulo $K$),\vspace{-0.2cm}
\end{itemize}
e.g., by moving $h_{i_k}$ just after $h_{i_{k+1}}$ in the demand preferences (without changing the ordering of other houses). We omit the mention of ``modulo $K$'' in the sequel.

Following standard notation, for any set $S\subseteq N$, $(\widetilde{\succeq}_{S}, \succeq_{-S})$ is the preference profile that is obtained from $\succeq$ when all agents $i\in S$ change their preferences from $\succeq_i$ to $\widetilde{\succeq}_{i}$. Consider the preference profile\vspace{-0.2cm}
$$\succeq^0=(\widetilde{\succeq}_{\{i_0,\ldots,i_{K}\}}, \succeq_{-\{i_0,\ldots,i_{K}\}}).\vspace{-0.2cm}$$
For each $0\leq k\leq K$, $\mathrm{TTC}_{i_k}(\succeq^0)=(h_{i_{k+1}},i_{k-1})$. Furthermore, by \textsl{individual rationality} of $\varphi$,  for each $0\leq k\leq K$, $\varphi_{i_k}(\succeq^0)\in \{(h_{i_k},i_k),(h_{i_{k+1}},\cdot)\}$. So, the set of houses allocated to agents in $\{i_0,\ldots,i_{K}\}$ at allocation $\varphi(\succeq^0)$ equals $\{h_{i_0},\ldots,h_{i_{K}}\}$. Hence, by \textsl{Pareto efficiency} of $\varphi$, for each $i_k\in \{i_0,\ldots,i_{K}\}$, $\varphi_{i_k}(\succeq^0)=(h_{i_{k+1}},i_{k-1}) =\mathrm{TTC}_{i_k}(\succeq^0)$.

Of course, original demand lexicographic preferences might not have been based on the specific demand preferences we just assumed; this is where \textsl{strategy-proofness} of $\varphi$ is used in an induction argument to change trading cycle agents' preferences one by one back to their original preferences, without changing the allotments of trading cycle agents under $\varphi$. To be more precise, we now use an induction argument on the number of agents who change their preferences starting from profile $\succeq^0$ and moving towards profile $\succeq$ to show that for each $i_k\in \{i_0,\ldots,i_{K}\}$, $\varphi_{i_k}(\succeq)=(h_{i_{k+1}},i_{k-1}) =\mathrm{TTC}_{i_k}(\succeq)$.\smallskip\pagebreak

\noindent \textbf{Induction 1 (for TTC algorithm Step~1 trading cycles)}\smallskip

\noindent \textit{Induction basis.} One agent $l_1$ changes his preferences starting from profile $\succeq^0$ and moving towards profile $\succeq$.

Let $l_1\in\{i_0,\ldots,i_{K}\}$ such that $l_1+1$ is the successor and $l_1-1$ is the predecessor in the trading cycle and $\mathrm{TTC}_{l_1}(\succeq^0)=(h_{l_1+1},l_1-1)$. Assume that starting from preference profile $\succeq^0$, agent $l_1$ changes his preferences from $\widetilde{\succeq}_{l_1}$ to $\succeq_{l_1}$, i.e., consider the preference profile\vspace{-0.2cm}
$$\succeq^1=(\widetilde{\succeq}_{\{i_0,\ldots,i_{K}\}\setminus\{l_1\}}, \succeq_{l_1},\succeq_{-\{i_0,\ldots,i_{K}\}}) =(\succeq_{l_1},\succeq^0_{-l_1}).\vspace{-0.2cm}$$
Then, since agent $l_1$'s trading cycle did not change, $\mathrm{TTC}_{l_1}\left(\succeq^0\right) = \mathrm{TTC}_{l_1}\left(\succeq^1\right)=(h_{l_1+1},l_1-1)$.  Considering the same preference change under rule $\varphi$, by \textsl{strategy-proofness} of $\varphi$, we have $\varphi_{l_1}(\succeq^1)\mathbin{\succeq_{l_1}}\varphi_{l_1}(\succeq^0)$.
Recall that at $\varphi_{l_1}(\succeq^0) =(h_{l_1+1},l_1-1)$, agent $l_1$ receives his favorite house according to demand preferences $\succeq_{l_1}^d$. Hence, $\varphi_{l_1}\left(\succeq^1\right)=(h_{l_1+1},\cdot)$. Next, by \textsl{individual rationality} of $\varphi$, for each $i_k\in \{i_0,\ldots,i_{K}\}\setminus \{l_1\}$, $\varphi_{i_k}\left(\succeq^1\right)\in \{(h_{i_k},i_k),(h_{i_{k+1}},\cdot)\}$. So, the set of houses allocated to agents in $\{i_0,\ldots,i_{K}\}$ at allocation $\varphi(\succeq^1)$ equals $\{h_{i_0},\ldots,h_{i_{K}}\}$. Then, by \textsl{Pareto efficiency} of $\varphi$, for each $i_k\in \{i_0,\ldots,i_{K}\}$, $\varphi_{i_k}(\succeq^1)=(h_{i_{k+1}},i_{k-1}) =\mathrm{TTC}_{i_k}(\succeq^1)$.\smallskip

\noindent \textit{Induction hypothesis.} The set of agents $\{l_1,\ldots,l_{\ell-1}\}\subseteq \{0,\ldots,K-1\}$ changes their preferences starting from profile $\succeq^0$ and moving towards profile $\succeq$. Then, for preference profile\vspace{-0.2cm}
$$\succeq^{\ell-1}=(\widetilde{\succeq}_{\{i_0,\ldots,i_{K}\}\setminus \{l_1,\ldots,l_{\ell-1}\}}, \succeq_{l_1},\ldots,\succeq_{l_{\ell-1}},\succeq_{-\{i_0,\ldots,i_{K}\}})\vspace{-0.2cm}$$
we have that for each $i_k\in \{i_0,\ldots,i_{K}\}$, $\varphi_{i_k}(\succeq^{\ell-1})=(h_{i_{k+1}},i_{k-1}) =\mathrm{TTC}_{i_k}(\succeq^{\ell-1})$.\smallskip

\noindent \textit{Induction step.} The set of agents $\{l_1,\ldots,l_{\ell}\}\subseteq \{0,\ldots,K-1\}$ changes their preferences starting from profile $\succeq^0$ and moving towards profile $\succeq$.

Let $l_{\ell}\in\{i_0,\ldots,i_{K}\}\setminus \{1,\ldots, l_{\ell-1}\}$ such that $l_{\ell}+1$ is the successor and $l_{\ell}-1$ is the predecessor in the trading cycle and $\mathrm{TTC}_{l_{\ell}}(\succeq^{\ell-1})=(h_{l_{\ell}+1},l_{\ell}-1)$. Assume that starting from preference profile $\succeq^{\ell-1}$, agent $l_{\ell}$ changes his preferences from $\widetilde{\succeq}_{l_{\ell}}$ to $\succeq_{l_{\ell}}$, i.e., consider the preference profile\vspace{-0.2cm}
$$\succeq^{\ell}=(\widetilde{\succeq}_{\{i_0,\ldots,i_{K}\}\setminus \{l_1,\ldots,l_{\ell}\}}, \succeq_{l_1},\ldots,\succeq_{l_{\ell}},\succeq_{-\{i_0,\ldots,i_{K}\}}) =(\succeq_{l_{\ell}},\succeq^{\ell-1}_{-l_{\ell}}).\vspace{-0.2cm}$$
Then, since agent $l_{\ell}$'s trading cycle did not change, $\mathrm{TTC}_{l_{\ell}}\left(\succeq^{\ell-1}\right) = \mathrm{TTC}_{l_{\ell}}\left(\succeq^{\ell}\right)=(h_{l_{\ell}+1},l_{\ell}-1)$.  Considering the same preference change under rule $\varphi$, by \textsl{strategy-proofness} of $\varphi$ we have $\varphi_{l_{\ell}}(\succeq^{\ell})\mathbin{\succeq_{l_{\ell}}} \varphi_{l_{\ell}}(\succeq^{\ell-1})$.
Recall that by the induction hypothesis, at $\varphi_{l_{\ell}}(\succeq^{\ell-1}) =(h_{l_{\ell}+1},l_{\ell}-1)$, agent $l_{\ell}$ receives his favorite house according to demand preferences $\succeq_{l_{\ell}}^d$. Hence,\vspace{-0.2cm}
\begin{itemize}
\item $\varphi_{l_{\ell}}\left(\succeq^{\ell}\right)=(h_{l_{\ell}+1},\cdot)$.\vspace{-0.2cm}
\end{itemize}

Since the choice of agent $l_{\ell}\in \{l_1,\ldots,l_{\ell}\}$ was arbitrary, we obtain by the same arguments as for agent $\ell$  that\vspace{-0.2cm}
\begin{itemize}
\item $\varphi_{l_{1}}\left(\succeq^{\ell}\right)=(h_{l_{1}+1},\cdot)$ [the induction hypothesis applies to the set of agents $\{l_1,\ldots,l_{\ell}\}\setminus\{l_1\}$],\vspace{-0.2cm}
\item[] \quad\quad\quad\quad\ $\vdots$\vspace{-0.2cm}
\item $\varphi_{l_{\ell-1}}\left(\succeq^{\ell}\right)=(h_{l_{\ell-1}+1},\cdot)$ [the induction hypothesis applies to the set of agents $\{l_1,\ldots,l_{\ell}\}\setminus\{l_{\ell-1}\}$].\vspace{-0.2cm}
\end{itemize}

Next, by \textsl{individual rationality} of $\varphi$, for each $i_k\in \{i_0,\ldots,i_{K}\}\setminus \{l_1,\ldots,l_{\ell}\}$, $\varphi_{i_k}\left(\succeq^{\ell}\right)\in \{(h_{i_k},i_k),(h_{i_{k+1}},\cdot)\}$. So, the set of houses allocated to agents in $\{i_0,\ldots,i_{K}\}$ at allocation $\varphi(\succeq^{\ell})$ equals $\{h_{i_0},\ldots,h_{i_{K}}\}$. Then, by \textsl{Pareto efficiency} of $\varphi$, for each $i_k\in \{i_0,\ldots,i_{K}\}$, $K\geq 1$, $\varphi_{i_k}(\succeq^{\ell})=(h_{i_{k+1}},i_{k-1}) =\mathrm{TTC}_{i_k}(\succeq^{\ell})$.

Thus, by \textsl{individual rationality} and Induction~$1$, we have shown that for each $i_k\in \{i_0,\ldots,i_{K}\}$, $K\geq 0$, $\varphi_{i_k}(\succeq) =\mathrm{TTC}_{i_k}(\succeq)$.

We have shown that agents who trade in the first step of the TTC algorithm always receive their TTC allotments under $\varphi$. Next, we consider agents who trade in the second step of the TTC. If such a trading cycle consists of only one agent, then that agent can never receive his most preferred house (since it is traded in Step~1) and, by \textsl{individual rationality} of $\varphi$, now receives his second most preferred own house. For larger trading cycles, we first consider demand preferences where trading cycle agents rank their ``TTC house'' first and their own house second and follow the same proof steps as for Step~1 trading cycle agents. Thus, the proof is done after an induction argument on the number of TTC algorithm steps.\smallskip

\noindent \textbf{Induction 2 on the number of TTC algorithm steps}\smallskip

\noindent \textit{Induction basis.} Consider a trading cycle that forms during Step~1 of the TTC algorithm for $\succeq$ with trading cycle agents $i_0,\ldots,i_{K}$, $K\geq 0$. Then, by \textsl{individual rationality} (for $K=0$) and Induction~1, we have shown that for each $i_k\in \{i_0,\ldots,i_{K}\}$, $\varphi_{i_k}(\succeq) ={\mathrm{TTC}_{i_k}(\succeq)}$.\smallskip

We assume that the TTC algorithm has at least $\kappa$ steps.\smallskip

\noindent \textit{Induction hypothesis.}  Consider a trading cycle that forms during Step~$\kappa -1$ of the TTC algorithm for $\succeq$ with trading cycle agents $i_0,\ldots,i_{K}$, $K\geq 0$. Then, for each $i_k\in \{i_0,\ldots,i_{K}\}$, $\varphi_{i_k}(\succeq)=\mathrm{TTC}_{i_k}(\succeq)$.\smallskip

\noindent \textit{Induction step.}  Consider a trading cycle that forms during Step~$\kappa$ of the TTC algorithm for $\succeq$. Recall that some houses have already been allocated to agents throughout Steps~$1,\ldots,{\kappa-1}$. We refer to these houses as \textit{allocated houses} and denote the \textit{set of allocated houses} by $\widetilde{H}$. By the induction hypothesis, the set of allocated houses does not depend on preferences of agents who receive their TTC allotments in Step~$\kappa$ and later steps.  We refer to the set of houses $H\setminus \widetilde{H}$ that are allocated in Step~$\kappa$ and later steps as \textit{available houses}.

If the trading cycle consists of only one agent, then that agent, according to his demand preferences, most prefers his own house among available houses $H\setminus \widetilde{H}$ and, by \textsl{individual rationality} of $\varphi$, receives it; such an agents receives his TTC allotment.

Hence, consider a Step~$\kappa$ trading cycle consisting of agents $i_0,\ldots,i_{K}$, $K\geq 1$, and houses $h_{i_0},\ldots,h_{i_{K}}$.  Note that each agent $i_k\in \{i_0,\ldots,i_{K}\}$, according to his demand preferences $\succeq_{i_k}^d$, prefers house $h_{i_{k+1}}$ most among available houses $H\setminus \widetilde{H}$.

For every $i_k\in \{i_0,\ldots,i_{K}\}$  we define preferences $\widetilde{\succeq}_{i_k}\,\in \mathcal{D}_{\mathrm{dlex}}$  based on new demand preferences $\widetilde{\succeq}^d$  and the original supply preferences ${\succeq}^s$ such that\vspace{-0.2cm}
\begin{itemize}
\item $h_{i_{k+1}}\mathbin{\widetilde{\succ}^d_{i_{k}}}h_{i_k} \mathbin{\widetilde{\succ}^d_{i_k}}\ldots$  (modulo $K$),\vspace{-0.2cm}
\end{itemize}
e.g., by moving $h_{i_{k+1}}$ in first and $h_{i_k}$ in second place in the demand preferences (without changing the ordering of other houses). We omit the mention of ``modulo $K$'' in the sequel.

Consider the preference profile \vspace{-0.2cm} $$\succeq^0=(\widetilde{\succeq}_{\{i_0,\ldots,i_{K}\}}, \succeq_{-\{i_0,\ldots,i_{K}\}}).\vspace{-0.2cm}$$
For each $0\leq k\leq K$, $\mathrm{TTC}_{i_k}(\succeq^0)=(h_{i_{k+1}},i_{k-1})$. Furthermore, by \textsl{individual rationality} of $\varphi$ for available houses $H\setminus \widetilde{H}$, for each $0\leq k\leq K$, $\varphi_{i_k}(\succeq^0)\in \{(h_{i_k},i_k),(h_{i_{k+1}},\cdot)\}$. So, the set of houses allocated to agents in $\{i_0,\ldots,i_{K}\}$ at allocation $\varphi(\succeq^0)$ equals $\{h_{i_0},\ldots,h_{i_{K}}\}$. Hence, by \textsl{Pareto efficiency} of $\varphi$, for each $i_k\in \{i_0,\ldots,i_{K}\}$, $\varphi_{i_k}(\succeq^0)=(h_{i_{k+1}},i_{k-1}) =\mathrm{TTC}_{i_k}(\succeq^0)$.

We now need to again use an induction argument on the number of agents who change their preferences starting from profile $\succeq^0$ and moving towards profile $\succeq$ to show that for each $i_k\in \{i_0,\ldots,i_{K}\}$, $\varphi_{i_k}(\succeq)=(h_{i_{k+1}},i_{k-1}) =\mathrm{TTC}_{i_k}(\succeq)$.\smallskip

\noindent \textbf{Induction $\bm{\kappa}$ (for TTC algorithm Step~$\bm{\kappa}$ trading cycles)}

The induction arguments for TTC algorithm Step~$\kappa$ trading cycles are essentially identical to those of Induction~1. The only difference is that instead of writing that ``an agent $i$ receives his favorite house according to his demand preferences $\succeq_{i}^d$'' we need to write that ``an agent $i$ receives his favorite \textit{available} house according to his demand preferences $\succeq_{i}^d$.'' Similarly, we use \textsl{individual rationality} of $\varphi$ for available houses $H\setminus \widetilde{H}$. We have shown that for each $i_k\in \{i_0,\ldots,i_{K}\}$, $\varphi_{i_k}(\succeq) =\mathrm{TTC}_{i_k}(\succeq)$.

Thus, by \textsl{individual rationality} and Induction~$\kappa$, we have shown that for each $i_k\in \{i_0,\ldots,i_{K}\}$, $K\geq 0$, $\varphi_{i_k}(\succeq) =\mathrm{TTC}_{i_k}(\succeq)$.

The Induction on the number of TTC algorithm steps is complete and it follows that $\varphi(\succeq) =\mathrm{TTC}(\succeq)$.
\end{proof}

\section{The domain of demand lexicographic preferences is not rich}\label{appendix:domainrichness}
The domain richness condition of \citet[][Assumption~B]{SonmezEco1999} for strict preferences over allotments reads as follows. For each preference relation $\succeq_i$ and each allotment $(h,j)\in\mathcal{A}_i$ such that $(h,j)\mathbin{\succ_i} (h_i,i)$, there exists a preference relation $\widetilde{\succeq}_i$ such that
\begin{itemize}
\item[(i)] the weak upper contour sets at $(h,j)$ under $\succeq_i$ and $\widetilde{\succeq}_i$ are the same, i.e., for all $(h',k)\in \mathcal{A}_i$, [$(h',k)\mathbin{\succeq_i}(h,j)$ if and only if $(h',k)\mathbin{\widetilde{\succeq}_i}(h,j)$],
\item[(ii)] the weak lower contour sets at $(h,j)$ under $\succeq_i$ and $\widetilde{\succeq}_i$ are the same, i.e., for all $(h',k)\in \mathcal{A}_i$, [$(h,j)\mathbin{\succeq_i}(h',k)$ if and only if $(h,j)\mathbin{\widetilde{\succeq}_i}(h',k)$], and
\item[(iii)] the endowment allotment ranks right after $(h,j)$, i.e., if $(h,j)\mathbin{\succ_i}(h',k)$, then $(h,j)\mathbin{\widetilde{\succ}_i}(h_i,i)\mathbin{\widetilde{\succeq}_i}(h',k)$.
\end{itemize}

To show that the domain of demand lexicographic preferences is not rich, consider the demand lexicographic preferences $\succeq_2$ for agent $2$ that we used in Example~\ref{example:multiSC}, $$(h_1,1)\mathbin{\succ_2}(h_1,3)\mathbin{\succ_2} \bm{(h_3,1)}\mathbin{\succ_2} (h_3,3)\mathbin{\succ_2} \bm{(h_2,2)}.$$
Then, \citeauthor{SonmezEco1999}'s domain richness condition would require that agent~2's demand lexicographic preferences $\succeq_2$ can be transformed either into preferences such that
$$(h_1,1)\mathbin{\widetilde{\succeq}_2}(h_1,3) \mathbin{\widetilde{\succeq}_2} \bm{(h_3,1)}\mathbin{\widetilde{\succ}_2} \bm{(h_2,2)}\mathbin{\widetilde{\succeq}_2} (h_3,3)\vspace{-0.2cm}$$ or into preferences such that
$$(h_1,3)\mathbin{\widetilde{\succeq}_2}(h_1,1) \mathbin{\widetilde{\succeq}_2} \bm{(h_3,1)}\mathbin{\widetilde{\succ}_2} \bm{(h_2,2)}\mathbin{\widetilde{\succeq}_2} (h_3,3).$$
However, neither of the latter preferences are demand lexicographic.

\end{appendix}
\end{document}